\documentclass[fleqn,english,authoryear]{elsarticle}
\usepackage[T1]{fontenc}
\usepackage[utf8]{inputenc}
\usepackage{geometry}
\usepackage{color}
\usepackage{babel}
\usepackage{array}
\usepackage{longtable}
\usepackage{float}
\usepackage{amsmath}
\usepackage{graphicx}
\usepackage[unicode=true,
 bookmarks=true,bookmarksnumbered=false,bookmarksopen=false,
 breaklinks=false,pdfborder={0 0 1},backref=false,colorlinks=false]{hyperref}
\usepackage{hyperref}
\hypersetup{colorlinks=true,citecolor=blue}

\makeatletter

\providecommand{\tabularnewline}{\\}

\usepackage{textcomp}
\usepackage{colortbl}
\usepackage{nccmath}
\usepackage{caption}
\usepackage{tabularx}

\let\ORIGtabular\tabular
\let\ORIGendtabular\endtabular
\let\ORIGtabularx\tabularx
\renewcommand*{\tabularx}{%
  \def\tabular{%
    \let\endtabular\ORIGendtabular
    \ORIGtabular
  }%
  \ORIGtabularx
}
\renewcommand{\tabular}{%
  \tabularx{\linewidth}%
}
\renewcommand{\endtabular}{\endtabularx}
\newcolumntype{Y}{>{\centering\arraybackslash}X}

\definecolor{lightgray}{gray}{0.9}

\@ifundefined{showcaptionsetup}{}{%
 \PassOptionsToPackage{caption=false}{subfig}}
\usepackage{subfig}
\makeatother

\begin{document}

\title{Managing connected and automated vehicles with flexible routing at
``lane-allocation-free'' intersections}

\author[author1]{Wanjing Ma}
\ead{mawanjing@tongji.edu.cn}
\author[author1,author2]{Ruochen Hao}
\ead{haoruochen@tongji.edu.cn}
\author[author1]{Chunhui Yu\corref{cor}}
\cortext[cor]{Corresponding author}
\ead{hughyu90@tongji.edu.cn}
\author[author1]{Tuo Sun}
\ead{suntuo77@163.com}
\author[author2]{Bart van Arem}
\ead{b.vanarem@tudelft.nl}

\address[author1]{Key Laboratory of Road and Traffic Engineering of the Ministry of
Education, Tongji University, 4800 Cao’an Road, Shanghai, P.R.China.}
\address[author2]{Department Transport and Planning, Faculty Civil Engineering and Geosciences, Delft University of Technology, Building 23, Stevinweg 1, 2628 CN, Delft, the Netherlands.}
\begin{abstract}
Trajectory planning and coordination for connected and automated vehicles
(CAVs) have been studied at isolated ``signal-free'' intersections
and in ``signal-free'' corridors under the fully CAV environment
in the literature. Most of the existing studies are based on the definition
of approaching and exit lanes. The route a vehicle takes to pass through
an intersection is determined from its movement. That is, only the
origin and destination arms are included. This study proposes a mixed-integer
linear programming (MILP) model to optimize vehicle trajectories at
an isolated ``signal-free'' intersection without lane allocation,
which is denoted as ``lane-allocation-free'' (LAF) control. Each
lane can be used as both approaching and exit lanes for all vehicle
movements including left-turn, through, and right-turn. A vehicle
can take a flexible route by way of multiple arms to pass through
the intersection. In this way, the spatial-temporal resources are
expected to be fully utilized. The interactions between vehicle trajectories
are modeled explicitly at the microscopic level. Vehicle routes and
trajectories (i.e., car-following and lane-changing behaviors) at
the intersection are optimized in one unified framework for system
optimality in terms of total vehicle delay. Considering varying traffic
conditions, the planning horizon is adaptively adjusted in the implementation
procedure of the proposed model to make a balance between solution
feasibility and computational burden. Numerical studies validate the
advantages of the proposed LAF control in terms of both vehicle delay
and throughput with different demand structures and temporal safety
gaps.
\end{abstract}
\maketitle

\section{Introduction}

With increasing traffic demand, vehicles suffer from severe traffic
congestion, which causes environmental problems and economic losses
\citep{Koonce2008}. Intersections are usually regarded as the bottlenecks
for traffic flows in an urban road network. Traffic management at
intersections is crucial to ensuring traffic efficiency, safety, energy
economics, and pollution reduction. Conventionally, priority rules
(e.g., stop signs, roundabouts, right-before-left, etc.) and traffic
signals are used to assign rights of way (ROW) to conflicting traffic
flows at an intersection. Fixed-time control, vehicle-actuated control,
and adaptive control are widely used in practice in terms of traffic
signal control \citep{Papageorgiou2003}. Numerous studies have been
dedicated to these research areas \citep{Allsop1976,Webster1958,Little1981,Heydecker1992,Han2014,Han2015,Liu2015,Memoli2017,Mohebifard2019,Mohajerpoor2019}.

With the popularity of connected and automated vehicles (CAVs), the
advances in CAV technologies are likely to produce a revolution in
traffic management \citep{Li2014a,pei2019cooperative}. The communications
between vehicles (V2V) and between vehicles and infrastructures (V2I)
can convey traffic information (e.g., signal timings, route guidance,
and speed advisory) from intersections to vehicles. At the same time,
detailed vehicle trajectory data (e.g., locations and speeds) can
be collected from vehicles for traffic management at intersections.
As CAVs are controllable, vehicle trajectory control becomes available
besides conventional signal control. It is expected that traffic control
would be implemented in both temporal and spatial dimensions.

A thorough review of the research on urban traffic signal control
with CAVs was provided in \citet{GUO2019313}. Generally, related
studies fall into three categories. In the first category, real-time
vehicle trajectory information (e.g., speeds and locations) is utilized
for signal optimization with or without infrastructure-based detector
data (e.g., traffic volumes from loop detectors) by catching real
time traffic demand (\citet{Gradinescu2007}) or temporal fluctuation
(\citet{feng2018better}). Signal timings such as cycle lengths and
green splits are optimized at isolated intersections \citep{Gradinescu2007,Guler2014,Feng2015,Liang2018,Yang2017}
and multiple intersections \citep{He2012,Yang2017}. The studies in
the second category focus on vehicle trajectory planning on the basis
of traffic information from intersections. One typical application
is eco-driving, which optimizes vehicle trajectories with the objectives
of minimizing fuel/energy consumption and emission. Typically, optimal
control models or feedback control models are formulated with vehicle
speeds or acceleration rates as the control variables \citep{Kamal2013,Wang2014,Wang2014a,Ubiergo2016,Wan2016}.
Platooning can also be considered (\citet{liu2019optimal,feng2019string}).
Approximation has then been proposed to solve the models more efficiently
by either discretizing time or segmenting trajectories \citep{Wan2016,Kamalanathsharma2013}.
In the third category, signal optimization and vehicle trajectory
planning are integrated into one unified framework. However, limited
studies have been reported. \citet{Li2014} enumerated feasible signal
plans and segmented vehicle trajectories for the joint optimization.
\citet{Feng2018} proposed a dynamic programming model for signal
optimization combined with an optimal control model for trajectory
planning as a two-stage model. \citet{Yu2018} proposed a mixed-integer
linear programming (MILP) model to simultaneously optimize signal
timings and vehicle trajectories. \citet{Guo2019} proposed a DP-SH
(dynamic programming with shooting heuristic) algorithm for efficiency
and jointly optimization of vehicle trajectories and signal timings.

Assuming the fully CAV environment, the concept of “signal-free” intersections
has been proposed \citep{Dresner2004,Dresner2008}. Vehicles cooperate
with each other and pass through intersections without physical traffic
signals. One prevailing category of such studies are based on the
philosophy of reservation. Approaching vehicles send requests to the
intersection controller to reserve space and time slots within the
intersection area. Reservation requests are managed to determine the
service sequence of the approaching vehicles, usually according to
rule-based policies such as ``first-come, first-served'' (FCFS)
strategy \citep{Au2010,Dresner2004,Dresner2008,Li2013}, priority
strategy \citep{Alonso2011}, auction strategy \citep{Carlino2013},
and platooning strategy \citep{Tachet2016}. However, both theoretical
analysis \citep{Yu2019a} and numerical case studies \citep{Levin2016}
showed that the advantages of reservation-based control might not
outperform conventional signal control (e.g., vehicle-actuated control)
in certain cases. Because the optimality cannot be guaranteed due
to the rule-based nature of reservation-based control. As a result,
optimization-based models have been proposed. In \citet{Bo2019Toward},
the scheduling algorithm can assign a feasible time to each arriving
AV with low complexity. Trajectory level optimization models are more
likely to take advantage of fully CAV environment. Typically, constrained
nonlinear optimization models are formulated \citep{Lee2012,Zohdy2016}.
In \citet{Lee2012}, vehicle acceleration/deceleration rates were
optimized to minimize trajectory overlap with the focus on safety.
In \citet{Zohdy2016}, vehicle arrival times at an intersection were
optimized to minimize vehicle delay with the focus on efficiency.
In addition, 3D CAV trajectories were mathematically formulated in
the combined temporal-spatial domains \citep{Li2019}. Priority-based
and Discrete Forward-Rolling Optimal Control (DFROC) algorithms were
developed for CAV management at isolated intersections. The optimization
of lane allocation is also considered in existing research, and proposed
CAV control Distributed control methods have also been investigated
to alleviate computational burden. \citet{Xu2018} projected approaching
vehicles from different traffic movements into a virtual lane and
then introduced a conflict-free geometry topology with the consideration
of the conflict relationship of involved vehicles. \citet{Mirheli2019}
proposed a vehicle-level mixed-integer non-linear programming model
for cooperative trajectory planning in a distributed way. Vehicle-level
solutions were pushed towards the global optimality.

Notwithstanding the abundant studies, it is noted that most of the
studies do not take into consideration the interactions of vehicle
trajectories at the microscopic level, which, however, is crucial
to vehicle trajectory planning. Car-following behaviors are usually
explicitly modeled while lane-changing behaviors are not. Recently,
\citet{Yu2019} successfully addressed this issue. Both car-following
and lane-changing behaviors of vehicles in a ``signal-free'' corridor
were cooperatively optimized in one unified framework. Approaching
lanes were not specified with lane allocation, which is called ``approaching-lane-allocation-free''
(ALAF) in this paper. Each approaching lane could be used by all vehicle
movements (i.e., left-turn, through, and right-turn). This study takes
a further step and eliminates the definition of approaching and exit
lanes. Acctually, it has been proved that the break of the traditional
division between approchang lanes and exiting lanes brings higher
control efficiency \citep{8889999,sun2017capacity}. However, the
lane allocations at intersections are fixed in these studies, and
the innovation which not divide traffic flows in directions is limited
in link scope. The dynamic lane allocation assignment is not considered.
In this study, the lane allocations of all lanes are elimited. Each
lane can be used by both approaching and leaving vehicles in all directions.
Further, the route a vehicle takes to pass through the intersection
is fixed in \citet{Yu2019}, which only consists of the origin and
the destination arms. In this study, a vehicle can take a flexible
route by way of multiple arms. In this way, the spatial-temporal resources
at intersections are expected to be fully utilized, especially with
imbalanced traffic. To this end, this study proposes an MILP model
to optimize vehicle routes and trajectories (i.e., car-following behaviors
and lane changing behaviors) at an isolated ``signal-free'' and
``lane-allocation-free'' intersection, which is denoted as ``lane-allocation-free''
(LAF) control. To balance solution feasibility and computational burden,
the planning horizon is adaptively adjusted in the implementation
procedure with varying traffic conditions.

The remainder of this paper is organized as follows. Section \ref{sec:Notations}
describes the problem and presents the notations. Section \ref{sec:Formulations}
formulates the MILP model to optimize vehicle routes and trajectories
at a ``signal-free'' and ``lane-allocation-free'' intersection.
Section \ref{sec:ImpleProcedure} presents the implementation procedure
of the proposed model with varying traffic conditions, which adaptively
adjusts the planning horizon to improve computational efficiency.
Numerical studies are conducted in Section \ref{sec:Numerical-studies}.Finally,
conclusions and recommendations are provided in Section \ref{sec:Conclusions}.

\section{Problem description and notations\label{sec:Notations}}

\subsection{Problem description}

\begin{figure}[tbph]
\noindent \begin{centering}
\includegraphics[width=0.8\linewidth]{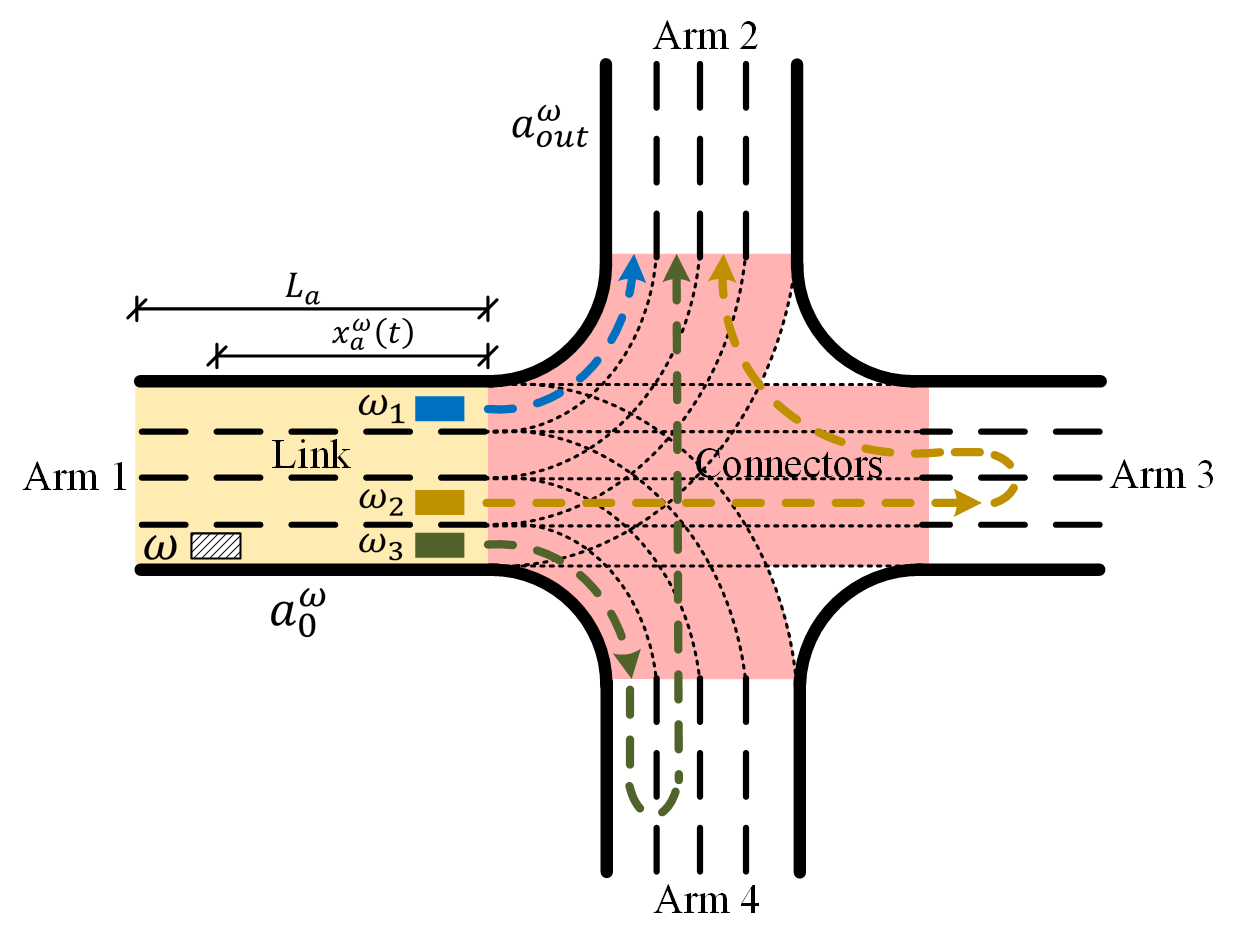}
\par\end{centering}
\caption{A ``signal-free'' and ``lane-allocation-free'' intersection with
four arms.\label{fig:IntersectionWithRouting}}
\end{figure}

Fig. \ref{fig:IntersectionWithRouting} shows a ``signal-free''
and ``lane-allocation-free'' intersection with four arms as an example.
In this study, one arm consists of an undirected link and all directed
connectors departing from the link. For example, arm 1 has four connectors
for left-turn traffic, four for through traffic, and four for right-turn
traffic. Fig. \ref{fig:IntersectionWithRouting} highlights the link
part and the connector part of arm 1. In contrast with conventional
intersections, no approaching lanes or exit lanes are defined, and
no lane allocation is specified. That is, each lane can be used by
both approaching and leaving vehicles in all directions in the control
zone at the intersection.

Conventionally, the route of a vehicle is fixed at an intersection.
For example, vehicle $\omega$ in arm 1 tries to turn left in Fig.
\ref{fig:IntersectionWithRouting}. It turns left directly similar
with the trajectory of vehicle $\omega_{1}$ under conventional traffic
management. That is, the route of vehicle $\omega$ only consists
of arm 1 and arm 2. If vehicle $\omega$ conflicts with other vehicles,
it may wait at the stop bar location, blocking the traffic behind.
Suppose there is heavy through traffic, light left-turn traffic in
arm 1 and light traffic in arm 4. Left-turn vehicle $\omega$ is waiting
in the rightmost lane in arm 1, looking for the gaps between the through
vehicles in the remaining three lanes. As a result, only three lanes
in arm 1 can be fully utilized at the same time. To improve the efficiency
of the intersection system, flexible routing is considered in this
study. Routing means that vehicle $\omega$ can travel to other arms
before entering the destination arm 2, e.g., following the trajectory
of vehicle $\omega_{3}$. In this way, vehicle $\omega$ can wait
in arm 4 instead of arm 1 and the four lanes in arm 1 can be fully
utilized by the heavy through traffic. The trajectory of vehicle $\omega_{2}$
is another possible route. In this way, it is expected the spatial-temporal
resources can be better utilized at the intersection.

Given the geometric layout of the intersection and the vehicles in
the control zone ($L_{a}$), the objective of this study is to cooperatively
optimize the routes and the trajectories of the vehicles for minimizing
total delay. The route plan of vehicle $\omega$ is the selection
of arms to be visited between the origin arm $a_{0}^{\omega}$, in
which vehicle $\omega$ is traveling, and the destination arm $a_{out}^{\omega}$
as well as the arm sequence. The trajectory of vehicle $\omega$ is
determined by the lane choice ($\delta_{k}^{\omega}(t)$) and the
longitudinal location ($x_{a}^{\omega}(t)$) in each visited arm at
each time step $t$. Note that $a_{0}^{\omega}$ is updated when vehicle
$\omega$ enters a new arm. For example, $a_{0}^{\omega}$ is arm
1 in Fig. \ref{fig:IntersectionWithRouting} and vehicle $\omega$
follows the trajectory of vehicle $\omega_{3}$. $a_{0}^{\omega}$
becomes arm 4 when vehicle $\omega$ travels in arm 4. $\mathbf{A}^{\omega}$
is then introduced to store the arms that vehicle $\omega$ has not
visited. In Fig. \ref{fig:IntersectionWithRouting}, $\mathbf{A}^{\omega}=\left\{ \mathrm{arm}\ 2,\mathrm{arm}\ 3,\mathrm{arm}\ 4\right\} $
when vehicle $\omega$ is in arm 1 and $\mathbf{A}^{\omega}=\left\{ \mathrm{arm}\ 2,\mathrm{arm}\ 3\right\} $
when vehicle $\omega$ travels into arm 4.

To simplify the formulations, the following assumptions are made:
\begin{itemize}
\item All vehicles are CAVs and can be controlled by a centralized controller.
\item The destination arm of a vehicle does not change after the vehicle
enters the control zone.
\item Vehicles follow the connectors and do not change lanes when traveling
within the intersection area.
\item Vehicles travel at constant speeds in connectors. The speed is determined
by the radius of a connector.
\item Vehicles can change lanes instantly in the link part of each arm.
\item Vehicle motion is captured by the first order model, the same assumption
as in Newell’s car-following model \citep{Newell2002}.
\end{itemize}

\subsection{Notations}

Main notations applied hereafter are summarized in Table \ref{tab:Notations}.
They will be explained in detail in the formulations.

\noindent %
\begin{longtable}[l]{l>{\raggedright}p{0.87\linewidth}}
\caption{Notations\label{tab:Notations}}
\tabularnewline
\endfirsthead
\hline 
\multicolumn{2}{l}{\textit{General notations}}\tabularnewline
\hline 
$\mathrm{M}$: & A sufficiently large number\tabularnewline
$t$: & Time step\tabularnewline
$\boldsymbol{\Omega}$: & Set of vehicles in the control zone of the intersection; each vehicle
is denoted as $\omega$\tabularnewline
$\mathbf{A}$: & Set of arms of the intersection; each arm is denoted as $a$\tabularnewline
$a_{0}^{\omega}$: & Origin arm in which vehicle $\omega$ is traveling when the optimization
is conducted\tabularnewline
$a_{out}^{\omega}$: & Destination arm in which vehicle $\omega$ leaves the control zone
of the intersection\tabularnewline
$\mathbf{A}^{\omega}$: & Set of arms that vehicle $\omega$ has not visited; if vehicle $\omega$
is in arm $a_{out}^{\omega}$, then $\mathbf{A}^{\omega}=\emptyset$\tabularnewline
$\mathbf{A}_{0}^{\omega}$: & Set of arms that vehicle $\omega$ is visiting or has not visited;
$\mathbf{A}_{0}^{\omega}=\mathbf{A}^{\omega}\cup\left\{ a_{0}^{\omega}\right\} $\tabularnewline
$\mathbf{K}_{a}$: & Set of lanes in arm $a$; each lane is denoted as $k$\tabularnewline
$\mathbf{K}_{a}^{left}$: & The leftmost lane of arm $a$\tabularnewline
$\mathbf{K}_{a}^{right}$: & The rightmost lane of arm $a$\tabularnewline
$k^{left}$: & The left adjacent lane of lane $k$ with facing the stop line\tabularnewline
$k^{right}$: & The right adjacent lane of lane $k$ with facing the stop line\tabularnewline
$\mathbf{K}_{a_{1}}^{a_{2}}$: & Set of lanes in arm $a_{1}$ that are connected to the lanes in arm
$a_{2}$\tabularnewline
$\mathbf{K}_{out}^{\omega}$: & Set of lanes in the destination arm in which vehicle $\omega$ leaves
the control zone\tabularnewline
$k_{+}^{a}$: & Succeeding lane of lane $k$ in arm $a$; that is, lane $k_{+}^{a}$
is connected from lane $k$ by a connector\tabularnewline
$\left\langle k_{1},k_{2}\right\rangle $: & Connector from lane $k_{1}$ to lane $k_{2}$\tabularnewline
$\mathbf{P}_{k_{1},k_{2}}^{k_{3},k_{4}}$: & Set of conflict points between connector $\left\langle k_{1},k_{2}\right\rangle $
and connector $\left\langle k_{3},k_{4}\right\rangle $; each conflict
point is denoted as $p$\tabularnewline
\hline 
\multicolumn{2}{l}{\textit{Parameters}}\tabularnewline
\hline 
$\Delta t$: & Length of time step, s\tabularnewline
$t_{0}$: & Current time when vehicle routes and trajectories are optimized, which
indicates the start of the planning horizon (i.e., $t=0$), s\tabularnewline
$T$: & Planning horizon; the horizon duration is $T\cdot\Delta t$\tabularnewline
$T_{0}$: & Initial value of $T$ in the implementation procedure for adaptively
adjusting $T$\tabularnewline
$\Delta T$: & Step length for adjusting $T$ in the implementation procedure\tabularnewline
$T^{turn}$: & Time steps of turnning around; the turnning around time is $T^{turn}\cdot\Delta t$\tabularnewline
$L_{a}$: & Length of the link part of arm $a$ in the control zone, m\tabularnewline
$V_{a}$: & Speed limit on the link part of arm $a$, m/s\tabularnewline
$l_{k_{1}}^{k_{2}}$: & Length of connector $\left\langle k_{1},k_{2}\right\rangle $ that
connects lane $k_{1}$and lane $k_{2}$, m\tabularnewline
$l_{k_{1},k_{2}}^{p}$: & Distance between the start of connector $\left\langle k_{1},k_{2}\right\rangle $
and conflict point $p$, m\tabularnewline
$v_{k_{1}}^{k_{2}}$: & Travel speed in connector $\left\langle k_{1},k_{2}\right\rangle $,
m/s\tabularnewline
$\tau$: & Temporal safety gap, s\tabularnewline
$d:$ & Spatial safety gap, m\tabularnewline
$\tilde{x}^{\omega}$: & Distance between vehicle $\omega$ and the stop bar location in the
current arm at the current time step, m\tabularnewline
$\tilde{\delta}_{k}^{\omega}$: & 1, if vehicle $\omega$ is in lane $k$ in the current arm at the
current time step; 0, otherwise\tabularnewline
$\tilde{dir}_{a}^{\omega}$: & 1, if vehicle $\omega$ driving toward the stop line; 0, otherwise\tabularnewline
$\tilde{\gamma}_{a_{1},a_{2}}^{\omega}$: & 1, if vehicle $\omega$ plans to travel from arm $a_{1}$ to arm $a_{2}$
according the previous optimization; 0, otherwise\tabularnewline
$\underline{t}_{0}^{\omega}$: & Recorded the time point when vehicle $\omega$ entered the link part
of the current arm, which is a relative value to the current time,
s\tabularnewline
$\bar{t}_{0}^{\omega}$: & Recorded the time point when vehicle $\omega$ left the link part
of the current arm into a connector, which is a relative value to
the current time, s\tabularnewline
$w_{1}/w_{2}$: & Weighting parameter in the objective function\tabularnewline
\hline 
\multicolumn{2}{l}{\textit{Decision variables}}\tabularnewline
\hline 
$x_{a}^{\omega}(t)$: & Distance from vehicle $\omega$ to the stop bar location in arm $a$
at time step $t$, m\tabularnewline
$\delta_{k}^{\omega}(t)$: & 1, if vehicle $\omega$ is in lane $k$ at time step $t$; 0, otherwise\tabularnewline
$\underline{t}_{a}^{\omega}/\bar{t}_{a}^{\omega}$: & Time point when vehicle $\omega$ enters/leaves the link part of arm
$a$, s\tabularnewline
$\gamma_{a_{1},a_{2}}^{\omega}$: & 1, if vehicle $\omega$ planes to travel from arm $a_{1}$ to arm
$a_{2}$ in the following time; 0, otherwise\tabularnewline
\hline 
\multicolumn{2}{l}{\textit{Auxiliary variables}}\tabularnewline
\hline 
$\underline{\mu}_{a}^{\omega}(t)$: & 1, if $t\cdot\Delta t\ge\underline{t}_{a}^{\omega}$; 0, otherwise\tabularnewline
$\bar{\mu}_{a}^{\omega}(t)$: & 1, if $t\cdot\Delta t\ge\bar{t}_{a}^{\omega}$; 0, otherwise\tabularnewline
$dir_{a}^{\omega}(t)$: & 1, if vehicle $\omega$ driving direction is toward the stop line
of arm $a$; 0, otherwise\tabularnewline
$ta_{a}^{\omega}(t)$: & 1, if vehicle $\omega$ is turning around at time $t$ in arm $a$;
0, otherwise\tabularnewline
$tal_{a}^{\omega}(t)$: & 1, if vehicle $\omega$ is turning around by using the left adjacent
lane at time $t$ in arm $a$; 0, otherwise\tabularnewline
$tar_{a}^{\omega}(t)$: & 1, if vehicle $\omega$ is turning around by using the right adjacent
lane at time $t$ in arm $a$; 0, otherwise\tabularnewline
$\beta_{a}^{\omega}$: & 1, if vehicle $\omega$ plans to visit arm $a$ in the following time;
0, otherwise\tabularnewline
$v_{a}^{\omega}$: & Travel speed within the intersection area after vehicle $\omega$
leaves arm $a$, m/s\tabularnewline
$\pi_{k_{1},k_{2}}^{\omega_{1},\omega_{2}}$: & 0, if vehicle $\omega_{1}$ enters connector $\left\langle k_{1},k_{2}\right\rangle $
after vehicle $\omega_{2}$ leaves connector $\left\langle k_{2},k_{1}\right\rangle $;
1, otherwise\tabularnewline
$\rho_{a}^{\omega_{1},\omega_{2}}(t)$: & 1, if vehicle $\omega_{1}$ and vehicle $\omega_{2}$ travel in the
same lane in the link part of arm $a$ at time step $t$; 0, otherwise\tabularnewline
\hline 
\end{longtable}

\section{Formulations\label{sec:Formulations}}

This section presents the MILP model based on discrete time to cooperatively
optimize the routes and the trajectories of the vehicles in the control
zone. The constraints and the objective function are presented in
the following sub-sections.

\subsection{Constraints}

Decision variable related constraints, vehicle motion related constraints,
and safety related constraints are introduced in this section. The
decision variables are constrained by variable domains and boundary
conditions at the start and end of the planning horizon. The vehicle
motion constraints deal with route planning, vehicle longitudinal
motion, and lance choices when entering or leaving the link part of
an arm. The safety constraints guarantee spatial-temporal safety gaps
between vehicles traveling in arms or within the intersection area.

\subsubsection{Domains of decision variables}

There are three types of decision variables for each vehicle $\omega$
in each arm $a$. $x_{a}^{\omega}(t)$ is the distance between vehicle
$\omega$ and the stop bar location in arm $a$ at time step $t$.
$x_{a}^{\omega}(t)$ is positive when vehicle $\omega$ is in the
link part of arm $a$. And $x_{a}^{\omega}(t)$ is negative when vehicle
$\omega$ is in the connector part of arm $a$. $\delta_{k}^{\omega}(t)$
indicates the lane choice of vehicle $\omega$. $\delta_{k}^{\omega}(t)=1$
if vehicle $\omega$ is in lane $k$ at time step $t$. $\underline{t}_{a}^{\omega}$
and $\bar{t}_{a}^{\omega}$ are the time points of entering and leaving
the link part of arm $a$, respectively. $\bar{t}_{a}^{\omega}$ is
the time of leaving the control zone if arm $a$ is the destination
arm $a_{out}^{\omega}$. $\underline{t}_{a}^{\omega}$ and $\bar{t}_{a}^{\omega}$
are continuous and they are relative values to the current time $t_{0}$.

Denote $a_{0}^{\omega}$ as the origin arm, in which vehicle $\omega$
is traveling. If vehicle $\omega$ is in the link part of arm $a_{0}^{\omega}$,
then $\underline{t}_{a}^{\omega}$ and $\bar{t}_{a}^{\omega}$ are
constrained by

\begin{equation}
\underline{t}_{a}^{\omega}=\underline{t}_{0}^{\omega}\le0,\forall a=a_{0}^{\omega};\omega\in\boldsymbol{\Omega}\label{eq:VarDomain_1}
\end{equation}

\begin{equation}
0\le\bar{t}_{a}^{\omega}\le T\cdot\Delta t,\forall a=a_{0}^{\omega};\omega\in\boldsymbol{\Omega}\label{eq:VarDemain_2}
\end{equation}

\noindent where $\underline{t}_{0}^{\omega}$ is the recorded time
point of entering the link of arm $a_{0}^{\omega}$, which is a relative
value to the current time $t_{0}$; $\boldsymbol{\Omega}$ is the
set of vehicles in the control zone. $\underline{t}_{a}^{\omega}$
is non-positive according to Eq. (\ref{eq:VarDomain_1}). If vehicle
$\omega$ is in the connector part of arm $a$, then the following
constraint of $\bar{t}_{a}^{\omega}$ will be added instead of Eq.
(\ref{eq:VarDemain_2}):

\begin{equation}
\bar{t}_{a}^{\omega}=\bar{t}_{0}^{\omega}\le0,\forall a=a_{0}^{\omega};\omega\in\bar{\boldsymbol{\Omega}}
\end{equation}

\noindent where $\bar{t}_{0}^{\omega}$ is the recorded time point
of leaving the link of arm $a$, which is a relative value to the
current time $t_{0}$.

For other arms (i.e., $a\ne a_{0}^{\omega}$), $\underline{t}_{a}^{\omega}$
and $\bar{t}_{a}^{\omega}$ are constrained by Eqs. (\ref{eq:VarDomain_4})–(\ref{eq:VarDomain_7}):

\begin{equation}
0\le\underline{t}_{a}^{\omega}\le T\cdot\Delta t+\mathrm{M}\left(1-\beta_{a}^{\omega}\right),\forall a\in\mathbf{A},a\ne a_{0}^{\omega};\omega\in\boldsymbol{\Omega}\label{eq:VarDomain_4}
\end{equation}

\begin{equation}
\underline{t}_{a}^{\omega}\le\bar{t}_{a}^{\omega}\le T\cdot\Delta t+\mathrm{M}\left(1-\beta_{a}^{\omega}\right),\forall a\in\mathbf{A},a\ne a_{0}^{\omega};\omega\in\boldsymbol{\Omega}\label{eq:VarDomain_5}
\end{equation}

\begin{equation}
-\mathrm{M}\beta_{a}^{\omega}\le\underline{t}_{a}^{\omega}-2T\cdot\Delta t\le\mathrm{M}\beta_{a}^{\omega},\forall a\in\mathbf{A},a\ne a_{0}^{\omega};\omega\in\boldsymbol{\Omega}\label{eq:VarDomain_6}
\end{equation}

\begin{equation}
-\mathrm{M}\beta_{a}^{\omega}\le\bar{t}_{a}^{\omega}-2T\cdot\Delta t\le\mathrm{M}\beta_{a}^{\omega},\forall a\in\mathbf{A},a\ne a_{0}^{\omega};\omega\in\boldsymbol{\Omega}\label{eq:VarDomain_7}
\end{equation}

\noindent where $\beta_{a}^{\omega}$ is an auxiliary binary variable.
$\beta_{a}^{\omega}=1$ if vehicle $\omega$ plans to visit arm $a$;$\beta_{a}^{\omega}=0$,
otherwise. If vehicle $\omega$ plans to visit arm $a$ (i.e., $\beta_{a}^{\omega}=1$),
Eqs. (\ref{eq:VarDomain_4}) and (\ref{eq:VarDomain_5}) will be effective.
Otherwise, Eqs. (\ref{eq:VarDomain_6}) and (\ref{eq:VarDomain_7})
will be effective. In that case, $\underline{t}_{a}^{\omega}$ and
$\bar{t}_{a}^{\omega}$ are set as $2T\cdot\Delta t$, which means
that vehicle $\omega$ will never enter arm $a$ in the planning horizon.

Before vehicle $\omega$ enters the link part of arm $a$, $x_{a}^{\omega}(t)$
is defined as zero:

\begin{equation}
-\mathrm{M}\underline{\mu}_{a}^{\omega}(t)\le x_{a}^{\omega}(t)\le\mathrm{M}\underline{\mu}_{a}^{\omega}(t),\forall t=0,\dots,T;a\in\mathbf{A};\omega\in\boldsymbol{\Omega}\label{eq:VarDomain_8}
\end{equation}

\noindent where $\underline{\mu}_{a}^{\omega}(t)$ is an auxiliary
binary variable. $\underline{\mu}_{a}^{\omega}(t)=1$ if vehicle $\omega$
has entered arm $a$ by time step $t$; $\underline{\mu}_{a}^{\omega}(t)=0$,
otherwise. Eq. (\ref{eq:VarDomain_8}) guarantees that $x_{a}^{\omega}(t)=0$
when $\underline{\mu}_{a}^{\omega}(t)=0$.

When vehicle $\omega$ travels in the link part of arm $a$, $x_{a}^{\omega}(t)$
is bounded by

\begin{equation}
-\mathrm{M}\left(1-\underline{\mu}_{a}^{\omega}(t)+\bar{\mu}_{a}^{\omega}(t)\right)\le x_{a}^{\omega}(t)\le L_{a},\forall t=0,\dots,T;a\in\mathbf{A}_{0}^{\omega};\omega\in\boldsymbol{\Omega}\label{eq:VarDomain_9}
\end{equation}

\noindent where $L_{a}$ is the length of the link part of arm $a$
within the control zone; $\bar{\mu}_{a}^{\omega}(t)$ is an auxiliary
binary variable. $\bar{\mu}_{a}^{\omega}(t)=1$ if vehicle $\omega$
has left the link part of arm $a$ by time step $t$; $\bar{\mu}_{a}^{\omega}(t)=0$,
otherwise. $\mathbf{A}_{0}^{\omega}$ is the set of arms that vehicle
$\omega$ is visiting or has not visited, which is updated when vehicle
$\omega$ enters an arm. Eq. (\ref{eq:VarDomain_9}) guarantees that
$0\le x_{a}^{\omega}(t)\le L_{a}$ when $\underline{\mu}_{a}^{\omega}(t)=1$
and $\bar{\mu}_{a}^{\omega}(t)=0$.

\begin{figure}[tbph]
\begin{centering}
\subfloat[\label{fig:non-destination arm}]{\begin{centering}
\includegraphics[width=0.5\linewidth]{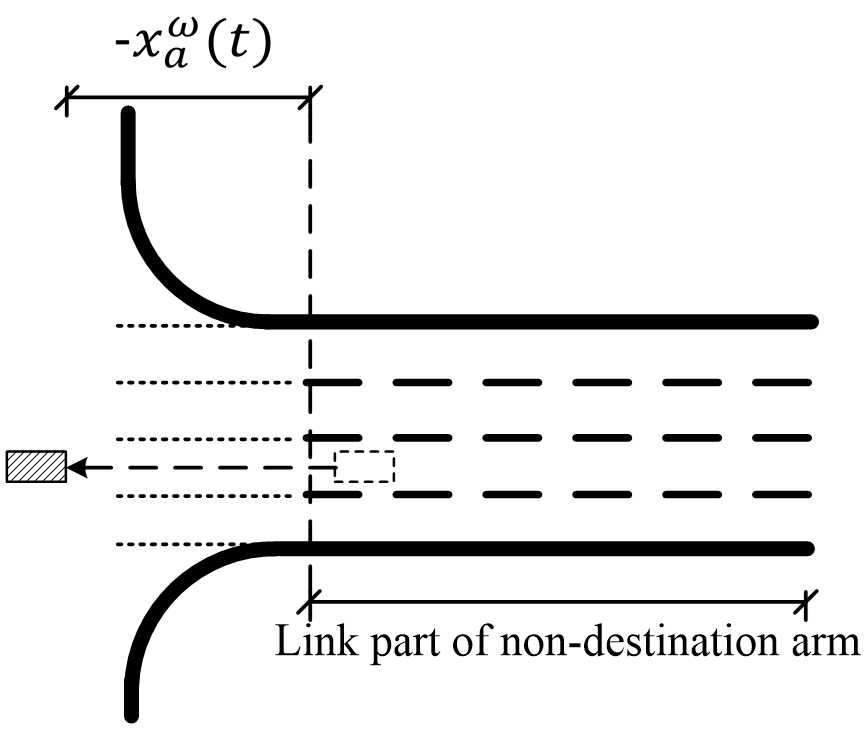}
\par\end{centering}
}\subfloat[\label{fig:destination arm}]{\begin{centering}
\includegraphics[width=0.5\linewidth]{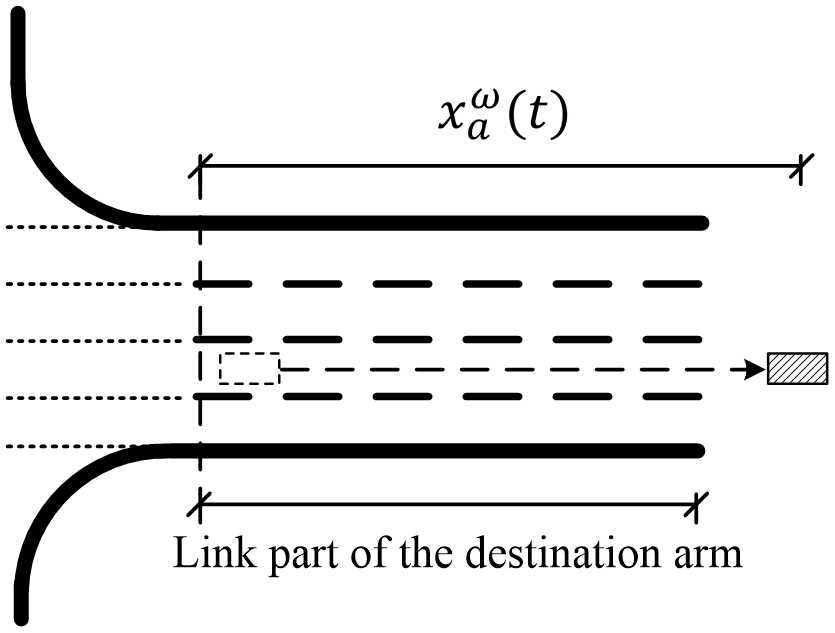}
\par\end{centering}
}
\par\end{centering}
\caption{Illustration of a vehicle leaving a link part: (a) In a non-destination
arm; (b) In a destination arm.}
\end{figure}

After vehicle $\omega$ leaves the link part of arm $a\ne a_{out}^{\omega}$
(i.e., $\bar{\mu}_{a}^{\omega}(t)=1$), $x_{a}^{\omega}(t)$ is defined
as a negative value, as shown in Fig.\ref{fig:non-destination arm}:

\begin{equation}
\begin{aligned} & -\mathrm{M}\left(1-\bar{\mu}_{a}^{\omega}(t)\right)\le x_{a}^{\omega}(t)+v_{a}^{\omega}\left(t\cdot\Delta t-\bar{t}_{a}^{\omega}\right)\le\mathrm{M}\left(1-\bar{\mu}_{a}^{\omega}(t)\right)\\
 & \forall t=0,\dots,T;a\in\mathbf{A}_{0}^{\omega},a\ne a_{out}^{\omega};\omega\in\boldsymbol{\Omega}
\end{aligned}
\label{eq:VarDomain_10}
\end{equation}

\noindent 
where $v_{a}^{\omega}$ is the travel speed of vehicle $\omega$ in
the connector part of arm $a$; $t\cdot\Delta t-\bar{t}_{a}^{\omega}$
is the travel time in the connector part at time step $t$. Eq. (\ref{eq:VarDomain_10})
indicates that $x_{a}^{\omega}(t)=-v_{a}^{\omega}\left(t\cdot\Delta t-\bar{t}_{a}^{\omega}\right)$
when $\bar{\mu}_{a}^{\omega}(t)=1$. $v_{a}^{\omega}$ is determined
by the planned route and the lane choice of vehicle $\omega$ when
leaving the link part of arm $a$:

\begin{equation}
\begin{aligned} & -\mathrm{M}\left(2-\gamma_{a_{1},a_{2}}^{\omega}-\delta_{k_{1}}^{\omega}\left(T\right)\right)\le v_{a_{1}}^{\omega}-v_{k_{1}}^{k_{2}}\le\mathrm{M}\left(2-\gamma_{a_{1},a_{2}}^{\omega}-\delta_{k_{1}}^{\omega}\left(T\right)\right)\\
 & \forall k_{2}=k_{1+}^{a_{2}};k_{1}\in\mathbf{K}_{a_{1}}^{a_{2}};a_{2}\in\mathbf{A}^{\omega},a_{2}\ne a_{1};a_{1}\in\mathbf{A}_{0}^{\omega},a_{1}\ne a_{out}^{\omega};\omega\in\boldsymbol{\Omega}
\end{aligned}
\label{eq:VarDomain_11}
\end{equation}

\noindent where $\mathbf{A}^{\omega}$ is the set of arms that have
not been visited, which is updated when vehicle $\omega$ enters an
arm; $\mathbf{K}_{a_{1}}^{a_{2}}$ is the set of lanes in arm $a_{1}$
that are connected to the lanes in arm $a_{2}$; $k_{1+}^{a_{2}}$
is the lane in arm $a_{2}$ that is connected from lane $k_{1}$ in
arm $a_{1}$; $v_{k_{1}}^{k_{2}}$ is the travel speed in connector
$\left\langle k_{1},k_{2}\right\rangle $. If vehicle $\omega$ travels
from lane $k_{1}$ in arm $a_{1}$ to lane $k_{2}$ in arm $a_{2}$
(i.e., $\gamma_{a_{1},a_{2}}^{\omega}=\delta_{k_{1}}^{\omega}\left(T\right)=1$),
Eq. (\ref{eq:VarDomain_11}) will set $v_{a_{1}}^{\omega}=v_{k_{1}}^{k_{2}}$.
Note that the final time step $T$ is used in Eq. (\ref{eq:VarDomain_11})
because $\delta_{k}^{\omega}\left(t\right)$ will be constrained to
remain the same after vehicle $\omega$ leaves the link part of an
arm.

After vehicle $\omega$ leaves the destination arm $a=a_{out}^{\omega}$
(i.e., $\bar{\mu}_{a}^{\omega}(t)=1$), $x_{a}^{\omega}(t)$ is set
as $L_{a}+V_{a}\left(t\cdot\Delta t-\bar{t}_{a}^{\omega}\right)$
as shown in Fig.\ref{fig:destination arm}:

\begin{equation}
\begin{aligned} & -\mathrm{M}\left(1-\bar{\mu}_{a}^{\omega}(t)\right)\le x_{a}^{\omega}(t)-\left(L_{a}+V_{a}\left(t\cdot\Delta t-\bar{t}_{a}^{\omega}\right)\right)\le\mathrm{M}\left(1-\bar{\mu}_{a}^{\omega}(t)\right)\\
 & \forall t=0,\dots,T;a=a_{out}^{\omega};\omega\in\boldsymbol{\Omega}
\end{aligned}
\label{eq:VarDomain_12}
\end{equation}

\noindent where $V_{a}$ is the speed limit on arm $a$; $t\cdot\Delta t-\bar{t}_{a}^{\omega}$
is the travel time in arm $a$ outside the control zone.

\subsubsection{Boundary conditions}

For the origin arm $a=a_{0}^{\omega}$ of vehicle $\omega$, $x_{a}^{\omega}(0)$
is determined by the current location of vehicle $\omega$:

\begin{equation}
x_{a}^{\omega}(0)=\tilde{x}^{\omega},\forall a=a_{0}^{\omega};\omega\in\boldsymbol{\Omega}
\end{equation}

\noindent where $\tilde{x}^{\omega}$ is the distance between vehicle
$\omega$ and the stop bar of the origin arm $a_{0}^{\omega}$ at
the current time. Similarly, the lane choice $\delta_{k}^{\omega}(0)$
in the origin arm $a=a_{0}^{\omega}$ is determined as well:

\begin{equation}
\delta_{k}^{\omega}(0)=\tilde{\delta}_{k}^{\omega},\forall k\in\mathbf{K}_{a};a=a_{0}^{\omega};\omega\in\boldsymbol{\Omega}
\end{equation}

\noindent where $\mathbf{K}_{a}$ is the set of lanes in arm $a$.
$\tilde{\delta}_{k}^{\omega}=1$ if vehicle $\omega$ is in lane $k$
at the current time; $\tilde{\delta}_{k}^{\omega}=0$, otherwise.
Apart from the initial lane and position, the initial driving direction
is also determined:

\begin{equation}
dir_{a}^{\omega}(0)=\tilde{dir}_{a}^{\omega},\forall a=a_{0}^{\omega};\omega\in\boldsymbol{\Omega}
\end{equation}
where $\tilde{dir}_{a}^{\omega}$ is the driving direction of vehicle
$\omega$ in the origin arm $a_{0}^{\omega}$ at the current time.
The driving directions in other arms are set as far from the stop
line:

\begin{equation}
dir_{a}^{\omega}(0)=1,\forall a\neq a_{0}^{\omega};\omega\in\boldsymbol{\Omega}
\end{equation}

At the end of the planning horizon, each vehicle $\omega$ is supposed
to have left the control zone of the intersection:

\begin{equation}
x_{a}^{\omega}(T)>L_{a},\forall a=a_{out}^{\omega};\omega\in\boldsymbol{\Omega}\label{eq:BoundaryCond_3}
\end{equation}

\subsubsection{Route planning}

$\gamma_{a_{1},a_{2}}^{\omega}$ is denoted as the indicator of the
arm sequence on the route of vehicle $\omega$. $\gamma_{a_{1},a_{2}}^{\omega}=1$
if vehicle $\omega$ plans to travel from arm $a_{1}$ to arm $a_{2}$;
$\gamma_{a_{1},a_{2}}^{\omega}=0$, otherwise. For the convenience
of modeling, $\gamma_{a_{1},a_{2}}^{\omega}$ is set as zero if $a_{1}=a_{2}$:

\begin{equation}
\gamma_{a,a}^{\omega}=0,\forall a\in\mathbf{A};\omega\in\boldsymbol{\Omega}\label{eq:RoutePlan_1}
\end{equation}

Each arm can be visited at most once by each vehicle, which is specified
by Eqs. (\ref{eq:RoutePlan_2})–(\ref{eq:RoutePlan_3}):

\begin{equation}
\sum_{a_{1}\in\mathbf{A}}\gamma_{a_{1},a_{2}}^{\omega}\le1,\forall a_{2}\in\mathbf{A};\omega\in\boldsymbol{\Omega}\label{eq:RoutePlan_2}
\end{equation}

\begin{equation}
\sum_{a_{2}\in\mathbf{A}}\gamma_{a_{1},a_{2}}^{\omega}\le1,\forall a_{1}\in\mathbf{A};\omega\in\boldsymbol{\Omega}\label{eq:RoutePlan_3}
\end{equation}

\noindent where $\sum_{a_{1}\in\mathbf{A}}\gamma_{a_{1},a_{2}}^{\omega}$
is the number of entering arms of arm $a_{2}$; $\sum_{a_{2}\in\mathbf{A}}\gamma_{a_{1},a_{2}}^{\omega}$
is the number of leaving arms of arm $a_{1}$.

If vehicle $\omega$ has visited arm $a$ (i.e., $a\in\mathbf{A}\setminus\mathbf{A}_{0}^{\omega}$),
then it will not visit this arm again, which is specified by Eqs.
(\ref{eq:RoutePlan_4}) and (\ref{eq:RoutePlan_5}):

\begin{equation}
\sum_{a_{1}\in\mathbf{A}}\gamma_{a_{1},a_{2}}^{\omega}=0,\forall a_{2}\in\mathbf{A}\setminus\mathbf{A}_{0}^{\omega};\omega\in\boldsymbol{\Omega}\label{eq:RoutePlan_4}
\end{equation}

\begin{equation}
\sum_{a_{2}\in\mathbf{A}}\gamma_{a_{1},a_{2}}^{\omega}=0,\forall a_{1}\in\mathbf{A}\setminus\mathbf{A}_{0}^{\omega};\omega\in\boldsymbol{\Omega}\label{eq:RoutePlan_5}
\end{equation}

Generally, there may be no connectors connecting arm $a_{1}$ and
arm $a_{2}$, e.g., because of forbidden vehicle movements. In that
case, $\mathbf{K}_{a_{1}}^{a_{2}}$ is an empty set. Then, $\gamma_{a_{1},a_{2}}^{\omega}$
should be zero if $\mathbf{K}_{a_{1}}^{a_{2}}$ is empty:

\begin{equation}
\gamma_{a_{1},a_{2}}^{\omega}\le\left|\mathbf{K}_{a_{1}}^{a_{2}}\right|,\forall a_{1},a_{2}\in\mathbf{A},a_{1}\ne a_{2};\omega\in\boldsymbol{\Omega}\label{eq:RoutePlan_6}
\end{equation}

\noindent where $\left|\mathbf{K}_{a_{1}}^{a_{2}}\right|$ is the
size of $\mathbf{K}_{a_{1}}^{a_{2}}$ (i.e., the number of the elements
in $\mathbf{K}_{a_{1}}^{a_{2}}$).

If the origin arm $a_{0}^{\omega}$ is not the destination arm $a_{out}^{\omega}$
(i.e., $a_{0}^{\omega}\ne a_{out}^{\omega}$), then vehicle $\omega$
will not enter arm $a_{0}^{\omega}$ from other arms in the following
time but leave arm $a_{0}^{\omega}$ to other arms:

\begin{equation}
\sum_{a_{1}\in\mathbf{A}}\gamma_{a_{1},a_{2}}^{\omega}=0,\forall a_{2}=a_{0}^{\omega},a_{2}\ne a_{out}^{\omega};\omega\in\boldsymbol{\Omega}\label{eq:RoutePlan_7}
\end{equation}

\begin{equation}
\sum_{a_{2}\in\mathbf{A}}\gamma_{a_{1},a_{2}}^{\omega}=1,\forall a_{1}=a_{0}^{\omega},a_{1}\ne a_{out}^{\omega};\omega\in\boldsymbol{\Omega}\label{eq:RoutePlan_8}
\end{equation}

If a non-destination arm $a_{1}$ is to be visited by vehicle $\omega$
(i.e., $a_{1}\in\mathbf{A}^{\omega},a_{1}\ne a_{out}^{\omega}$),
then the number of entering arms of arm $a_{1}$ should be equal to
the number of leaving arm of arm $a_{1}$, which are both one or zero:

\begin{equation}
\sum_{a_{2}\in\mathbf{A}}\gamma_{a_{1},a_{2}}^{\omega}=\sum_{a_{2}\in\mathbf{A}}\gamma_{a_{2},a_{1}}^{\omega},\forall a_{1}\in\mathbf{A}^{\omega},a_{1}\ne a_{out}^{\omega};\omega\in\boldsymbol{\Omega}\label{eq:RoutePlan_9}
\end{equation}

If the destination arm $a_{out}^{\omega}$ is not the origin one (i.e.,
$a_{out}^{\omega}\ne a_{0}^{\omega}$), then vehicle $\omega$ will
enter the arm $a_{out}^{\omega}$ from other arms in the following
planning horizon but bot leave arm $a_{out}^{\omega}$ to other arms:

\begin{equation}
\sum_{a_{1}\in\mathbf{A}}\gamma_{a_{1},a_{2}}^{\omega}=1,\forall a_{2}=a_{out}^{\omega},a_{2}\ne a_{0}^{\omega};\omega\in\boldsymbol{\Omega}\label{eq:RoutePlan_10}
\end{equation}

\begin{equation}
\sum_{a_{2}\in\mathbf{A}}\gamma_{a_{1},a_{2}}^{\omega}=0,\forall a_{1}=a_{out}^{\omega},a_{1}\ne a_{0}^{\omega};\omega\in\boldsymbol{\Omega}\label{eq:RoutePlan_11}
\end{equation}

If the destination arm $a_{out}^{\omega}$ is the origin one (i.e.,
$a_{out}^{\omega}=a_{0}^{\omega}$), then vehicle $\omega$ will not
travel from other arms to arm $a_{out}^{\omega}$ or from arm $a_{out}^{\omega}$
to other arms. It only travels in the destination arm until it leaves
the control zone.

\begin{equation}
\sum_{a_{1}\in\mathbf{A}}\gamma_{a_{1},a_{2}}^{\omega}=\sum_{a_{1}\in\mathbf{A}}\gamma_{a_{2},a_{1}}^{\omega}=0,\forall a_{2}=a_{out}^{\omega},a_{2}=a_{0}^{\omega};\omega\in\boldsymbol{\Omega}\label{eq:RoutePlan_12}
\end{equation}

If vehicle $\omega$ is traveling in the connector part of the origin
arm $a_{0}^{\omega}$, vehicle $\omega$ will not change lanes within
the intersection area. That is, the succeeding arm remains the same:

\begin{equation}
\gamma_{a_{1},a_{2}}^{\omega}=\tilde{\gamma}_{a_{1},a_{2}}^{\omega},\forall a_{2}\in\mathbf{A};a_{1}=a_{0}^{\omega};\omega\in\bar{\boldsymbol{\Omega}}\label{eq:RoutePlan_13}
\end{equation}

\noindent where $\tilde{\gamma}_{a_{1},a_{2}}^{\omega}$ indicates
the route planned in the previous optimization.

$\beta_{a_{1}}^{\omega}$ is introduced to indicate whether vehicle
$\omega$ plans to visit arm $a_{1}$ in the following time. If so,
$\beta_{a_{1}}^{\omega}=1$; otherwise, $\beta_{a_{1}}^{\omega}=0$.
This is guaranteed by Eqs. (\ref{eq:RoutePlan_14}) to (\ref{eq:RoutePlan_16}).

\begin{equation}
-\left(\sum_{a_{2}\in\mathbf{A}}\gamma_{a_{1},a_{2}}^{\omega}+\sum_{a_{2}\in\mathbf{A}}\gamma_{a_{2},a_{1}}^{\omega}\right)\le\beta_{a_{1}}^{\omega}\le\sum_{a_{2}\in\mathbf{A}}\gamma_{a_{1},a_{2}}^{\omega}+\sum_{a_{2}\in\mathbf{A}}\gamma_{a_{2},a_{1}}^{\omega},\forall a_{1}\in\mathbf{A};\omega\in\boldsymbol{\Omega}\label{eq:RoutePlan_14}
\end{equation}

\begin{equation}
\sum_{a_{2}\in\mathbf{A}}\gamma_{a_{1},a_{2}}^{\omega}\le\beta_{a_{1}}^{\omega},\forall a_{1}\in\mathbf{A};\omega\in\boldsymbol{\Omega}\label{eq:RoutePlan_15}
\end{equation}

\begin{equation}
\sum_{a_{1}\in\mathbf{A}}\gamma_{a_{1},a_{2}}^{\omega}\le\beta_{a_{2}}^{\omega},\forall a_{2}\in\mathbf{A};\omega\in\boldsymbol{\Omega}\label{eq:RoutePlan_16}
\end{equation}

\noindent If vehicle $\omega$ does not plan to visit arm $a_{1}$
(i.e., $\sum_{a_{2}\in\mathbf{A}}\gamma_{a_{1},a_{2}}^{\omega}=\sum_{a_{2}\in\mathbf{A}}\gamma_{a_{2},a_{1}}^{\omega}=0$),
then Eq. (\ref{eq:RoutePlan_14}) guarantees that $\beta_{a_{1}}^{\omega}=0$.
Otherwise, Eqs. (\ref{eq:RoutePlan_15}) and (\ref{eq:RoutePlan_16})
guarantee that $\beta_{a_{1}}^{\omega}=1$. Eqs. (\ref{eq:RoutePlan_15})
and (\ref{eq:RoutePlan_16}) are both set in case of the origin arm
and the destination arm. Because vehicle $\omega$ neither travels
from other arms into the origin arm nor travels from the destination
arm to other arms.

\subsubsection{Vehicle longitudinal motion}

\begin{figure}[tbph]
\noindent \begin{centering}
\subfloat[\label{fig:VehDynamics_a}]{\noindent \begin{centering}
\includegraphics[width=0.5\linewidth]{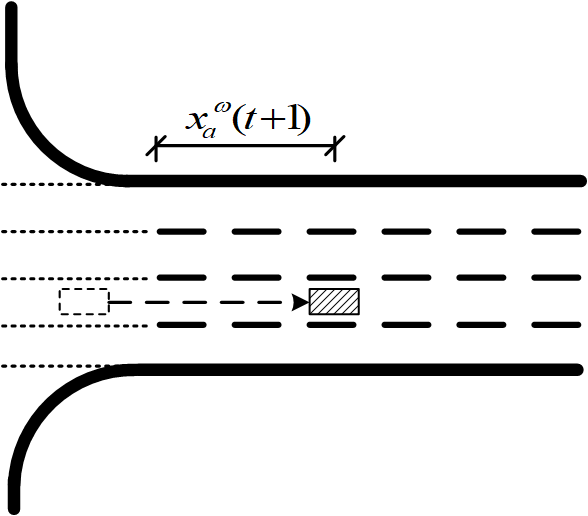}
\par\end{centering}
}\subfloat[\label{fig:VehDynamics_b}]{\noindent \begin{centering}
\includegraphics[width=0.5\linewidth]{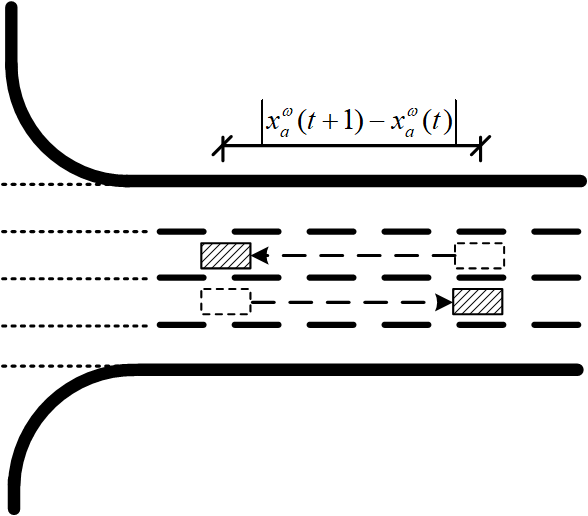}
\par\end{centering}
}
\par\end{centering}
\noindent \begin{centering}
\subfloat[\label{fig:VehDynamics_c}]{\noindent \begin{centering}
\includegraphics[width=0.5\linewidth]{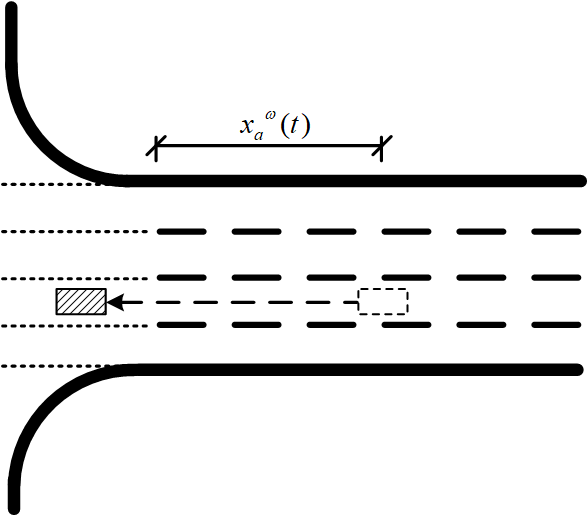}
\par\end{centering}
}\subfloat[\label{fig:VehDynamics_d}]{\noindent \begin{centering}
\includegraphics[width=0.5\linewidth]{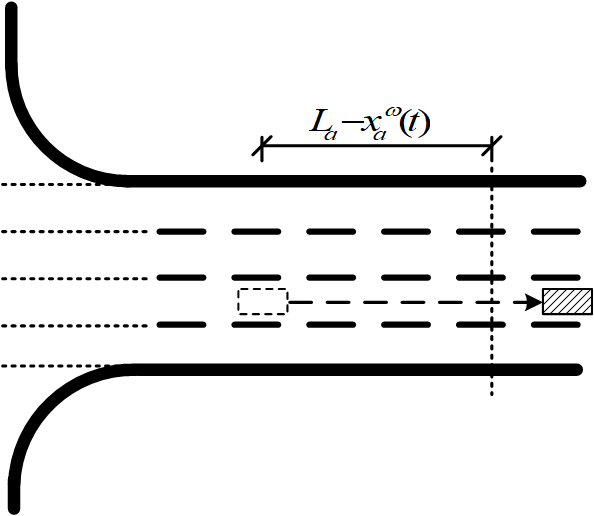}
\par\end{centering}
}
\par\end{centering}
\caption{Illustration of vehicle movements: (a) Enter the link part of an arm;
(b) Travel in the link part of an arm; (c) Leave the link part of
a non-destination arm; and (d) Leave the destination arm.}
\end{figure}

If vehicle $\omega$ enters the link part of arm $a$ during time
step $t+1$ (i.e., $\underline{\mu}_{a}^{\omega}(t)=0$ and $\underline{\mu}_{a}^{\omega}(t+1)=1$)
as shown in Fig. \ref{fig:VehDynamics_a}, the traveled distance in
the link part during this time step is constrained by the speed limit
$V_{a}$ on arm $a$:

\begin{equation}
\begin{aligned} & x_{a}^{\omega}(t+1)\le V_{a}\left(\left(t+1\right)\Delta t-\underline{t}_{a}^{\omega}\right)+\mathrm{M}\left(1+\underline{\mu}_{a}^{\omega}(t)-\underline{\mu}_{a}^{\omega}(t+1)\right)\\
 & \forall t=0,\dots,T-1;a\in\mathbf{A}_{0}^{\omega};\omega\in\boldsymbol{\Omega}
\end{aligned}
\label{eq:VehDynamics_1}
\end{equation}

\noindent where $\underline{\mu}_{a}^{\omega}(t)$ is an auxiliary
variable. $\underline{\mu}_{a}^{\omega}(t)=1$ if vehicle $\omega$
has entered the link part of arm $a$ by time step $t$; $\underline{\mu}_{a}^{\omega}(t)=0$,
otherwise. $\left(t+1\right)\Delta t-\underline{t}_{a}^{\omega}$
is the travel time in arm $a$ within time step $t+1$.

If vehicle $\omega$ travels in the link part of arm $a$ during time
step $t+1$ (i.e., $\underline{\mu}_{a}^{\omega}(t)=1$ and $\bar{\mu}_{a}^{\omega}(t+1)=0$)
as shown in Fig. \ref{fig:VehDynamics_b}, there will besimilar constraints:

\begin{equation}
\begin{aligned} & \left|x_{a}^{\omega}(t+1)-x_{a}^{\omega}(t)\right|\le V_{a}\Delta t+\mathrm{M}\left(1-\underline{\mu}_{a}^{\omega}(t)+\bar{\mu}_{a}^{\omega}(t+1)\right)\\
 & \forall t=0,\dots,T-1;a\in\mathbf{A}_{0}^{\omega};\omega\in\boldsymbol{\Omega}
\end{aligned}
\label{eq:VehDynamics_2}
\end{equation}

\noindent where $\bar{\mu}_{a}^{\omega}(t)$ is an auxiliary variable.
$\bar{\mu}_{a}^{\omega}(t)=1$ if vehicle $\omega$ has left the link
part of arm $a$ by time step $t$; $\bar{\mu}_{a}^{\omega}(t)=0$,
otherwise. Since vehicle $\omega$ can move both directions, the absolute
value function is used in Eq. (\ref{eq:VehDynamics_2}).

If vehicle $\omega$ leaves the link part of a non-destination arm
$a\ne a_{out}^{\omega}$ during time step $t+1$ (i.e., $\bar{\mu}_{a}^{\omega}(t)=0$
and $\bar{\mu}_{a}^{\omega}(t+1)=1$) as shown in Fig. \ref{fig:VehDynamics_c},
there will be:

\begin{equation}
\begin{aligned} & x_{a}^{\omega}(t)\le V_{a}\left(\bar{t}_{a}^{\omega}-t\cdot\Delta t\right)+\mathrm{M}\left(1+\bar{\mu}_{a}^{\omega}(t)-\bar{\mu}_{a}^{\omega}(t+1)\right)\\
 & \forall t=0,\dots,T-1;a\in\mathbf{A}_{0}^{\omega},a\ne a_{out}^{\omega};\omega\in\boldsymbol{\Omega}
\end{aligned}
\label{eq:VehDynamics_3}
\end{equation}

\noindent where $\bar{t}_{a}^{\omega}-t\cdot\Delta t$ is the travel
time in the link part of arm $a$ before vehicle $\omega$ leaves
the link part within time step $t+1$.

If vehicle $\omega$ leaves the control zone in the destination arm
$a_{out}^{\omega}$ during time step $t+1$ (i.e., $\bar{\mu}_{a}^{\omega}(t)=0$
and $\bar{\mu}_{a}^{\omega}(t+1)=1$) as shown in Fig. \ref{fig:VehDynamics_d},
there will be:

\begin{equation}
\begin{alignedat}{1} & L_{a}-x_{a}^{\omega}(t)\le V_{a}\left(\bar{t}_{a}^{\omega}-t\cdot\Delta t\right)+\mathrm{M}\left(1+\bar{\mu}_{a}^{\omega}(t)-\bar{\mu}_{a}^{\omega}(t+1)\right)\\
 & \forall t=0,\dots,T-1;a=a_{out}^{\omega};\omega\in\boldsymbol{\Omega}
\end{alignedat}
\label{eq:VehDynamics_4}
\end{equation}

\noindent different from Eq. (\ref{eq:VehDynamics_2}), $L_{a}$ is
used in Eq. (\ref{eq:VehDynamics_4}). Because vehicle $\omega$ leaves
the control zone in the destination arm.

\noindent 

The relationship between the driving direction and the longitudinal
position are constrained by Eqs. (\ref{eq:VehDynamics_5-1}) and (\ref{eq:VehDynamics_5-2}).

\begin{equation}
\begin{aligned} & x_{a}^{\omega}(t+1)-x_{a}^{\omega}(t)\le Mdir_{a}^{\omega}(t)\\
 & \forall t=0,\dots,T-1;\omega\in\boldsymbol{\Omega};a\in\mathbf{A}
\end{aligned}
\label{eq:VehDynamics_5-1}
\end{equation}

\begin{equation}
\begin{aligned} & x_{a}^{\omega}(t)-x_{a}^{\omega}(t+1)\le M\left(1-dir_{a}^{\omega}(t)\right)\\
 & \forall t=0,\dots,T-1;\omega\in\boldsymbol{\Omega};a\in\mathbf{A}
\end{aligned}
\label{eq:VehDynamics_5-2}
\end{equation}
where $dir_{a}^{\omega}(t)$ is the driving direction of vehicle $\omega$
in arm $a$ at time $t$. If $dir_{a}^{\omega}(t)=0$, $x_{a}^{\omega}(t+1)$
will not be larger than $x_{a}^{\omega}(t)$, which means vehicles
will always get close to the stop line, as shown as $\omega_{1}$
in \ref{fig:Illstration-of-left}.

A vehicle will stay idling if it is turnning around. This is guaranteed
by Eq. (\ref{eq:VehDynamics_6-2}).

\begin{equation}
\begin{aligned} & -M\left(1-ta_{a}^{\omega}(t-t^{turn})\right)\le x_{a}^{\omega}(t)-x_{a}^{\omega}(t+1)\le M\left(1-ta_{a}^{\omega}(t-t^{turn})\right)\\
 & \forall t=0,\dots,T-1;t^{turn}=0,...,T^{turn};\omega\in\boldsymbol{\Omega};a\in\mathbf{A}
\end{aligned}
\label{eq:VehDynamics_6-2}
\end{equation}
where $ta_{a}^{\omega}(t)$ is an auxiliary variable. $ta_{a}^{\omega}(t)=1$
if vehicle $\omega$ is turning around in arm $a$ at time $t$; $ta_{a}^{\omega}(t)=0$,
otherwise. $T^{turn}$is the turning around time.

The driving direction has to change after turning around:

\begin{equation}
\begin{aligned} & 1-M\left(1-ta_{a}^{\omega}(t)\right)\leq dir_{a}^{\omega}(t)+dir_{a}^{\omega}(t+1)\le1+M\left(1-ta_{a}^{\omega}(t)\right)\\
 & \forall t=0,\dots,T-1;\omega\in\boldsymbol{\Omega};a\in\mathbf{A}
\end{aligned}
\label{eq:VehDynamics_7-1}
\end{equation}
Eq. (\ref{eq:LaneChoice_7-1}) indicates that $dir_{a}^{\omega}(t)+dir_{a}^{\omega}(t+1)=1$
if $ta_{a}^{\omega}(t)=1$, which means the driving directions of
vehicle $\omega$ at time step $t$ and time step $t+1$ will be different
if vehicle $\omega$ turns around at time step $t$.

On the contrary, vehicles cannot change the driving direction without
turning around:

\begin{equation}
\begin{aligned} & dir_{a}^{\omega}(t)-Mta_{a}^{\omega}(t)\leq dir_{a}^{\omega}(t+1)\le dir_{a}^{\omega}(t)+Mta_{a}^{\omega}(t)\\
 & \forall t=0,\dots,T-1;\omega\in\boldsymbol{\Omega};a\in\mathbf{A}
\end{aligned}
\label{eq:VehDynamics_8-1}
\end{equation}

Considering the comfortable and rationality of vehicle driving, vehicles
can not turn around instantly:

\begin{equation}
\begin{alignedat}{1} & -(1+ta_{a}^{\omega}(t-1)-ta_{a}^{\omega}(t))\leq ta_{a}^{\omega}(t+t^{turn})\leq1+ta_{a}^{\omega}(t-1)-ta_{a}^{\omega}(t),\\
 & \forall t=1,\dots,T-T^{turn};t^{turn}=0,...,T^{turn};\omega\in\boldsymbol{\Omega};a\in\mathbf{A}
\end{alignedat}
\label{eq:42}
\end{equation}
where $T^{turn}$ is the time steps of turnning around.

\subsubsection{Lane choices}

At any time step in the planning horizon, vehicle $\omega$ can only
occupy one lane except when vehicle $\omega$ is turning around:

\begin{equation}
\begin{aligned} & 1+\sum_{t^{turn}=0}^{T^{turn}}ta_{a}^{\omega}\left(t-t^{turn}\right)-(1-\beta_{a}^{\omega})M\le\sum_{k\in\mathbf{K}_{a}}\delta_{k}^{\omega}\left(t\right)\\
 & \le1+\sum_{t^{turn}=0}^{T^{turn}}ta_{a}^{\omega}\left(t-t^{turn}\right)+(1-\beta_{a}^{\omega})M,\forall t=0,\dots,T;a\in\mathbf{A};\omega\in\boldsymbol{\Omega}
\end{aligned}
\label{eq:LaneChoice_1}
\end{equation}

\begin{equation}
-\beta_{a}^{\omega}M\le\sum_{k\in\mathbf{K}_{a}}\delta_{k}^{\omega}\left(t\right)\le\beta_{a}^{\omega}M,\forall t=0,\dots,T;t^{turn}=0,...,T^{turn};a\in\mathbf{A};\omega\in\boldsymbol{\Omega}\label{eq:LaneChoice_2}
\end{equation}

\noindent if vehicle $\omega$ plans to visit arm $a$ in the following
time (i.e., $\beta_{a}^{\omega}=1$), then Eq. (\ref{eq:LaneChoice_1})
will be effective and $\sum_{k\in\mathbf{K}_{a}}\delta_{k}^{\omega}\left(t\right)=1+\sum_{t^{turn}=0}^{T^{turn}}ta_{a}^{\omega}\left(t-t^{turn}\right)$.
According to the constrain Eq. (\ref{eq:42}), the gap between two
turning around is larger than $T^{turn}$, which means $\sum_{t^{turn}=0}^{T^{turn}}ta_{a}^{\omega}\left(t-t^{turn}\right)$
will equal to 1 only if the vehicle turns around in last $T^{turn}$
seconds. Otherwise, Eq. (\ref{eq:LaneChoice_2}) is effective and
$\sum_{k\in\mathbf{K}_{a}}\delta_{k}^{\omega}\left(t\right)=0$.

It is assumed that vehicle $\omega$ can only change one lane within
one time step. That is, if vehicle $\omega$ is in lane $k_{1}$ at
time step $t$ (i.e., $\delta_{k_{1}}^{\omega}\left(t\right)=1$),
then it can only take its current or adjacent lanes at time step $t+1$:

\begin{equation}
\begin{aligned} & \delta_{k_{1}}^{\omega}\left(t\right)-1\le\delta_{k_{2}}^{\omega}\left(t+1\right)\le1-\delta_{k_{1}}^{\omega}\left(t\right)\\
 & \forall t=0,\dots,T-1;k_{1},k_{2}\in\mathbf{K}_{a},\left|k_{2}-k_{1}\right|\ge2;a\in\mathbf{A}_{0}^{\omega};\omega\in\boldsymbol{\Omega}
\end{aligned}
\label{eq:LaneChoice_3}
\end{equation}

\noindent Eq. (\ref{eq:LaneChoice_3}) sets $\delta_{k_{2}}^{\omega}\left(t+1\right)=0$
when $\delta_{k_{1}}^{\omega}\left(t\right)=0$ and $\left|k_{2}-k_{1}\right|\ge2$.
That is, vehicle $\omega$ cannot change more than one lanes within
one time step.

If vehicle $\omega$ is idling during time step $t+1$ (i.e., $x_{a}^{\omega}(t)=x_{a}^{\omega}(t+1)$),
then it cannot change lanes and should remain in its current lane
(i.e., $\delta_{k}^{\omega}\left(t\right)=\delta_{k}^{\omega}\left(t+1\right)$):

\begin{equation}
\begin{alignedat}{1} & -\mathrm{M}\left(x_{a}^{\omega}(t)-x_{a}^{\omega}(t+1)\right)\le\delta_{k}^{\omega}\left(t\right)-\delta_{k}^{\omega}\left(t+1\right)\le\mathrm{M}\left(x_{a}^{\omega}(t)-x_{a}^{\omega}(t+1)\right)\\
 & \forall t=0,\dots,T-1;k\in\mathbf{K}_{a};a\in\mathbf{A}_{0}^{\omega};\omega\in\boldsymbol{\Omega}
\end{alignedat}
\label{eq:LaneChoice_4}
\end{equation}

To avoid blocking incoming vehicles, the lane in which vehicle $\omega$
leaves the control zone is constrained as:

\begin{equation}
\bar{\mu}_{a}^{\omega}(t)-1\le\sum_{k\in\mathbf{K}_{out}^{\omega}}\delta_{k}^{\omega}\left(t\right)-1\le1-\bar{\mu}_{a}^{\omega}(t),\forall t=0,\dots,T;a=a_{out}^{\omega};\omega\in\boldsymbol{\Omega}\label{eq:LaneChoice_5}
\end{equation}

\noindent where $\mathbf{K}_{out}^{\omega}$ is the set of lanes those
vehicle $\omega$ can use to leave the control zone in the destination
arm $a_{out}^{\omega}$. Eq. (\ref{eq:LaneChoice_5}) guarantees that
vehicle $\omega$ leaves the control zone in one lane of $\mathbf{K}_{out}^{\omega}$
(i.e., $\sum_{k\in\mathbf{K}_{out}^{\omega}}\delta_{k}^{\omega}\left(t\right)=1$
when $\bar{\mu}_{a}^{\omega}(t)=1$).

\begin{figure}[tbph]
\begin{centering}
\textcolor{red}{\includegraphics[width=0.5\linewidth]{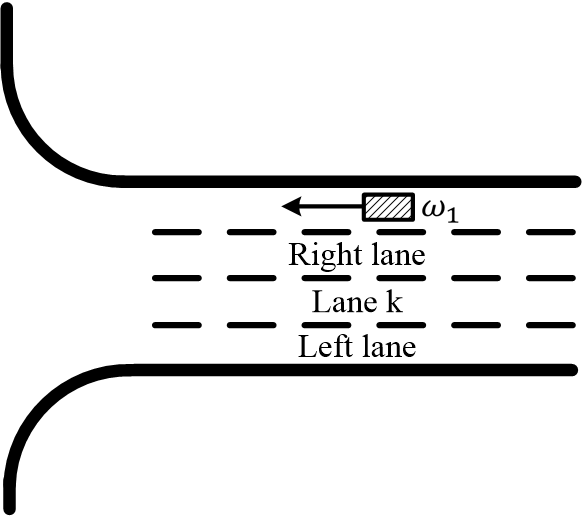}}
\par\end{centering}
\textcolor{red}{\caption{Illstration of left adjacent lane and lane number\label{fig:Illstration-of-left}}
}
\end{figure}

Two lanes need to be occupied during the process of turning around.
When vehicle $\omega$ turns around from its left side, the lane and
its left adjacent lane of driving direction need to be occupied. They
are realized by Eqs. (\ref{eq:LaneChoice_6-1}) and (\ref{eq:LaneChoice_6-2}).

\begin{equation}
\begin{aligned} & 2-M\left(3-\delta_{k^{left}}^{\omega}(t-t^{turn})-\delta_{k}^{\omega}(t-t^{turn})-tal_{a}^{\omega}(t-t^{turn})+dir_{a}^{\omega}(t)\right)\leq\delta_{k^{left}}^{\omega}(t)+\delta_{k}^{\omega}(t)\\
 & \le2+M\left(3-\delta_{k^{left}}^{\omega}(t-t^{turn})-\delta_{k}^{\omega}(t-t^{turn})-tal_{a}^{\omega}(t-t^{turn})+dir_{a}^{\omega}(t)\right)\\
 & \forall t=0,\dots,T-1;\omega\in\boldsymbol{\Omega};k\neq\mathbf{K}_{a}^{left};t^{turn}=0,...,T^{turn};a\in\mathbf{A}
\end{aligned}
\label{eq:LaneChoice_6-1}
\end{equation}

\begin{equation}
\begin{aligned} & 2-M\left(4-\delta_{k^{right}}^{\omega}(t-t^{turn})-\delta_{k}^{\omega}(t-t^{turn})-tal_{a}^{\omega}(t-t^{turn})-dir_{a}^{\omega}(t)\right)\leq\delta_{k^{right}}^{\omega}(t)+\delta_{k}^{\omega}(t)\\
 & \le2+M\left(4-\delta_{k^{right}}^{\omega}(t-t^{turn})-\delta_{k}^{\omega}(t-t^{turn})-tal_{a}^{\omega}(t-t^{turn})-dir_{a}^{\omega}(t)\right)\\
 & \forall t=0,\dots,T-1;\omega\in\boldsymbol{\Omega};k\neq\mathbf{K}_{a}^{right};t^{turn}=0,...,T^{turn};a\in\mathbf{A}
\end{aligned}
\label{eq:LaneChoice_6-2}
\end{equation}
where $tal_{a}^{\omega}(t)$ is an auxiliary variable. $tal_{a}^{\omega}(t)=1$
if vehicle $\omega$ turns around from its left side in arm $a$ at
time $t$; $tal_{a}^{\omega}(t)=0$, otherwise. $k^{left}$ is the
left adjacent lane of lane $k$ with facing the stop line as shown
in Fig. \ref{fig:Illstration-of-left}. Similarly, $k^{right}$ is
the right adjacent lane of lane $k$ with facing the stop line. $\mathbf{K}_{a}^{left}$
indicates the leftmost lane of the link part of arm $a$, while $\mathbf{K}_{a}^{right}$
indicates the rightmost lane of the link part of arm $a$.

\textcolor{red}{}

Similarly, the lane and the right adjacent lane of driving direction
need to be occupied when a vehicle turns around from its right side:

\begin{equation}
\begin{aligned} & 2-M\left(3-\delta_{k^{right}}^{\omega}(t-t^{turn})-\delta_{k}^{\omega}(t-t^{turn})-tar_{a}^{\omega}(t-t^{turn})+dir_{a}^{\omega}(t)\right)\leq\delta_{k^{right}}^{\omega}(t)+\delta_{k}^{\omega}(t)\\
 & \le2+M\left(3-\delta_{k^{right}}^{\omega}(t-t^{turn})-\delta_{k}^{\omega}(t-t^{turn})-tar_{a}^{\omega}(t-t^{turn})+dir_{a}^{\omega}(t)\right)\\
 & \forall t=0,\dots,T-1;\omega\in\boldsymbol{\Omega};k\neq\mathbf{K}_{a}^{right};t^{turn}=0,...,T^{turn};a\in\mathbf{A}
\end{aligned}
\label{eq:LaneChoice_7-1}
\end{equation}

\begin{equation}
\begin{aligned} & 2-M\left(4-\delta_{k^{left}}^{\omega}(t-t^{turn})-\delta_{k}^{\omega}(t-t^{turn})-tar_{a}^{\omega}(t-t^{turn})-dir_{a}^{\omega}(t)\right)\leq\delta_{k^{left}}^{\omega}(t)+\delta_{k}^{\omega}(t)\\
 & \le2+M\left(4-\delta_{k^{left}}^{\omega}(t-t^{turn})-\delta_{k}^{\omega}(t-t^{turn})-tar_{a}^{\omega}(t-t^{turn})-dir_{a}^{\omega}(t)\right)\\
 & \forall t=0,\dots,T-1;\omega\in\boldsymbol{\Omega};k\neq\mathbf{K}_{a}^{left};t^{turn}=0,...,T^{turn};a\in\mathbf{A}
\end{aligned}
\label{eq:LaneChoice_7-2}
\end{equation}
where $tar_{a}^{\omega}(t)$ is an auxiliary variable. $tar_{a}^{\omega}(t)=1$
if vehicle $\omega$ turns around from its right side in arm $a$
at time $t$; $tar_{a}^{\omega}(t)=0$, otherwise.

There is no doubt that vehicles have to turn around from either its
left side or its right side:

\begin{equation}
\begin{aligned} & tar_{a}^{\omega}(t)+tal_{a}^{\omega}(t)=ta{}_{a}^{\omega}(t)\\
 & \forall t=0,\dots,T;\omega\in\boldsymbol{\Omega};a\in\mathbf{A}
\end{aligned}
\label{eq:LaneChoice_8-1}
\end{equation}
where $ta_{a}^{\omega}(t)$ is an auxiliary variable. $ta_{a}^{\omega}(t)=1$
if vehicle $\omega$ turns around in arm $a$ at time $t$; $ta_{a}^{\omega}(t)=0$,
otherwise.

\begin{figure}[tbph]
\begin{centering}
\textcolor{red}{\includegraphics[width=0.5\linewidth]{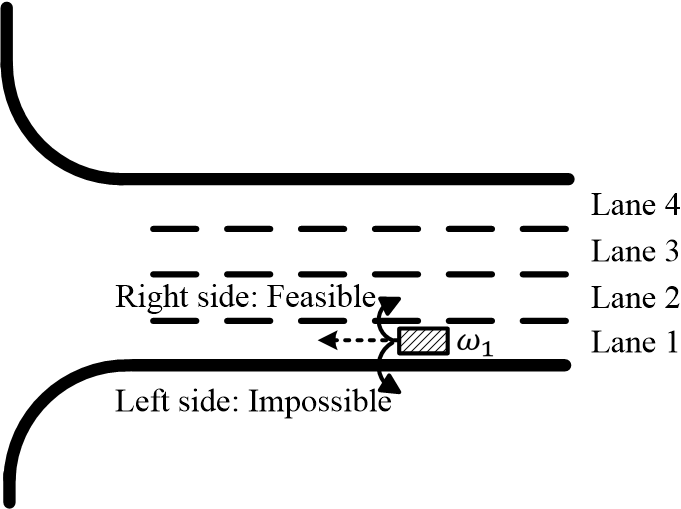}}
\par\end{centering}
\textcolor{red}{\caption{Illstration that vehicle can not turn afound from the left side\label{fig:Illstration-that-vehicle}}
}
\end{figure}

However, sometimes vehicles can not turn around from its left or its
right side. For instance, a vehicle can not turn around from the left
side when it drives in the leftmost lane of the arm, as shown as $\omega_{1}$
in Fig. \ref{fig:Illstration-that-vehicle}. This is guaranteed by:

\begin{equation}
\begin{aligned} & -M\left(1-\delta_{k}^{\omega}(t)+dir_{a}^{\omega}(t)\right)\leq tal_{a}^{\omega}(t)\le M\left(1-\delta_{k}^{\omega}(t)+dir_{a}^{\omega}(t)\right)\\
 & \forall t=0,\dots,T;\omega\in\boldsymbol{\Omega};k=\mathbf{K}_{a}^{left};a\in\mathbf{A}
\end{aligned}
\label{eq:LaneChoice_9-1}
\end{equation}

\begin{equation}
\begin{aligned} & -M\left(2-\delta_{k}^{\omega}(t)-dir_{a}^{\omega}(t)\right)\leq tal_{a}^{\omega}(t)\le M\left(2-\delta_{k}^{\omega}(t)-dir_{a}^{\omega}(t)\right)\\
 & \forall t=0,\dots,T;\omega\in\boldsymbol{\Omega};k=\mathbf{K}_{a}^{right};a\in\mathbf{A}
\end{aligned}
\label{eq:LaneChoice_9-2}
\end{equation}
\begin{equation}
\begin{aligned} & -M\left(1-\delta_{k}^{\omega}(t)+dir_{a}^{\omega}(t)\right)\leq tar_{a}^{\omega}(t)\le M\left(1-\delta_{k}^{\omega}(t)+dir_{a}^{\omega}(t)\right)\\
 & \forall t=0,\dots,T;\omega\in\boldsymbol{\Omega};k=\mathbf{K}_{a}^{right};a\in\mathbf{A}
\end{aligned}
\label{eq:LaneChoice_10-1}
\end{equation}

\begin{equation}
\begin{aligned} & -M\left(2-\delta_{k}^{\omega}(t)-dir_{a}^{\omega}(t)\right)\leq tar_{a}^{\omega}(t)\le M\left(2-\delta_{k}^{\omega}(t)-dir_{a}^{\omega}(t)\right)\\
 & \forall t=0,\dots,T;\omega\in\boldsymbol{\Omega};k=\mathbf{K}_{a}^{left};a\in\mathbf{A}
\end{aligned}
\label{eq:LaneChoice_10-2}
\end{equation}
Eqs. (\ref{eq:LaneChoice_9-1}) and (\ref{eq:LaneChoice_9-2}) indicate
the situation that vehicle $\omega$ can not turn around from the
left side, while Eqs. (\ref{eq:LaneChoice_10-1}) and (\ref{eq:LaneChoice_10-2})
indicate the situation that vehicle $\omega$ can not turn around
from the right side.

\textcolor{red}{}

\subsubsection{Entering an arm}

\begin{figure}[tbph]
\begin{centering}
\includegraphics[width=0.6\linewidth]{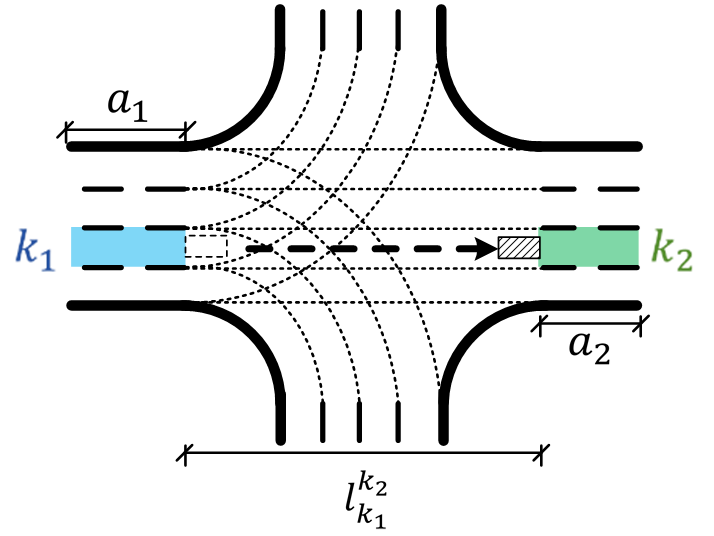}
\par\end{centering}
\caption{Illustration of a vehicle entering an arm.\label{fig:Illustration-of-vehicles}}
\end{figure}

If vehicle $\omega$ plans to travel from arm $a_{1}$ to arm $a_{2}$
(i.e., $\gamma_{a_{1},a_{2}}^{\omega}=1$), then the entering time
$\underline{t}_{a_{2}}^{\omega}$ will be determined by the leaving
time $\bar{t}_{a_{1}}^{\omega}$ and the travel time in the connectors
within the intersection area as shown in Fig.\ref{fig:Illustration-of-vehicles}:

\begin{equation}
\begin{alignedat}{1} & -\mathrm{M}\left(2-\gamma_{a_{1},a_{2}}^{\omega}-\delta_{k_{1}}^{\omega}\left(T\right)\right)\le\underline{t}_{a_{2}}^{\omega}-\left(\bar{t}_{a_{1}}^{\omega}+\frac{l_{k_{1}}^{k_{2}}}{v_{k_{1}}^{k_{2}}}\right)\le\mathrm{M}\left(2-\gamma_{a_{1},a_{2}}^{\omega}-\delta_{k_{1}}^{\omega}\left(T\right)\right)\\
 & \forall k_{2}=k_{1+}^{a_{2}};k_{1}\in\mathbf{K}_{a_{1}}^{a_{2}};a_{1},a_{2}\in\mathbf{A}_{0}^{\omega};\omega\in\boldsymbol{\Omega}
\end{alignedat}
\label{eq:EnterArm_1}
\end{equation}
The last time step $T$ is used in Eq. (\ref{eq:EnterArm_1}) to indicate
the lane in which vehicle $\omega$ leaves, the same as Eq. (\ref{eq:VarDomain_11}).
$l_{k_{1}}^{k_{2}}$ and $v_{k_{1}}^{k_{2}}$ are the length and travel
speed in connector $\left\langle k_{1},k_{2}\right\rangle $.

\noindent 

Further, the lane in which vehicle $\omega$ enters arm $a_{2}$ is
determined by the lane in which vehicle $\omega$ leaves the link
part of arm $a_{1}$:

\begin{equation}
\gamma_{a_{1},a_{2}}^{\omega}-1\le\delta_{k_{2}}^{\omega}\left(0\right)-\delta_{k_{1}}^{\omega}\left(T\right)\le1-\gamma_{a_{1},a_{2}}^{\omega},\forall k_{2}=k_{1+}^{a_{2}};k_{1}\in\mathbf{K}_{a_{1}}^{a_{2}};a_{1},a_{2}\in\mathbf{A}_{0}^{\omega};\omega\in\boldsymbol{\Omega}\label{eq:EnterArm_2}
\end{equation}

\noindent Eq. (\ref{eq:EnterArm_2}) indicates that vehicle $\omega$
will enter arm $a_{2}$ in lane $k_{2}$ if it leaves the link part
of arm $a_{1}$ in lane $k_{1}$ (i.e., $\delta_{k_{2}}^{\omega}\left(0\right)=\delta_{k_{1}}^{\omega}\left(T\right)$
when $\gamma_{a_{1},a_{2}}^{\omega}=1$).

Vehicles are not permitted to change lanes when entering an arm. If
vehicle $\omega$ enters arm $a_{2}$ during time step $t+1$ (i.e.,
$\underline{\mu}_{a}^{\omega}(t)=0$ and $\underline{\mu}_{a}^{\omega}(t+1)=1$),
its lane choice should keep the same (i.e., $\delta_{k}^{\omega}\left(t\right)=\delta_{k}^{\omega}\left(t+1\right)$):

\begin{equation}
\begin{alignedat}{1} & -\left(1+\underline{\mu}_{a}^{\omega}(t)-\underline{\mu}_{a}^{\omega}(t+1)\right)\le\delta_{k}^{\omega}\left(t\right)-\delta_{k}^{\omega}\left(t+1\right)\le1+\underline{\mu}_{a}^{\omega}(t)-\underline{\mu}_{a}^{\omega}(t+1)\\
 & \forall t=0,\dots,T-1;k\in\mathbf{K}_{a};a\in\mathbf{A}^{\omega};\omega\in\boldsymbol{\Omega}
\end{alignedat}
\label{eq:EnterArm_3}
\end{equation}

$\underline{\mu}_{a}^{\omega}(t)$ is an auxiliary variable for the
convenience of modeling. It is related to $\underline{t}_{a}^{\omega}$
in the following Eq. (\ref{eq:EnterArm_4}):

\begin{equation}
-\mathrm{M}\left(1-\underline{\mu}_{a}^{\omega}(t)\right)\le t\cdot\Delta t-\underline{t}_{a}^{\omega}\le\mathrm{M}\underline{\mu}_{a}^{\omega}(t),\forall t=0,\dots,T;a\in\mathbf{A};\omega\in\boldsymbol{\Omega}\label{eq:EnterArm_4}
\end{equation}

\noindent Eq. (\ref{eq:EnterArm_4}) indicates that $\underline{\mu}_{a}^{\omega}(t)=1$
if $t\cdot\Delta t\ge\underline{t}_{a}^{\omega}$; $\underline{\mu}_{a}^{\omega}(t)=0$,
otherwise.

\subsubsection{Leaving the link part of an arm}

If vehicle $\omega$ leaves the link part of a non-destination arm
$a\ne a_{out}^{\omega}$ during time step $t+1$ (i.e., $\bar{\mu}_{a}^{\omega}(t)=0$
and $\bar{\mu}_{a}^{\omega}(t+1)=1$) as shown in Fig. \ref{fig:VehDynamics_c},
then $x_{a}^{\omega}(t)\ge0$ and $x_{a}^{\omega}(t+1)<0$. The above
Eqs. (\ref{eq:VarDomain_8}) and (\ref{eq:VarDomain_9}) guarantee
that $x_{a}^{\omega}(t)\ge0$ when $\bar{\mu}_{a}^{\omega}(t)=0$.
Eq. (\ref{eq:VarDomain_10}) guarantees that $x_{a}^{\omega}(t+1)<0$
when $\bar{\mu}_{a}^{\omega}(t+1)=1$. If vehicle $\omega$ leaves
the link part of the destination arm $a_{out}^{\omega}$ during time
step $t+1$ as shown in Fig. \ref{fig:VehDynamics_d}, then $L_{a}\ge x_{a}^{\omega}(t)\ge0$
and $x_{a}^{\omega}(t+1)\ge L_{a}$ will be guaranteed by Eqs. (\ref{eq:VarDomain_8}),
(\ref{eq:VarDomain_9}), and (\ref{eq:VarDomain_12}).

When vehicle $\omega$ leaves the link part of a non-destination arm
$a_{1}\ne a_{out}^{\omega}$, the selected lane is constrained as:

\begin{equation}
\gamma_{a_{1},a_{2}}^{\omega}-1\le\sum_{k\in\mathbf{K}_{a_{1}}^{a_{2}}}\delta_{k}^{\omega}(T)-1\le1-\gamma_{a_{1},a_{2}}^{\omega},\forall a_{1},a_{2}\in\mathbf{A}_{0}^{\omega},a_{1}\ne a_{out}^{\omega};\omega\in\boldsymbol{\Omega}\label{eq:LeaveArm_1}
\end{equation}

\noindent If vehicle $\omega$ plans to travel from arm $a_{1}\ne a_{out}^{\omega}$
to arm $a_{2}$ (i.e., $\gamma_{a_{1},a_{2}}^{\omega}=1$), then one
lane in $\mathbf{K}_{a_{1}}^{a_{2}}$ will be used. On the other hand,
$\gamma_{a_{1},a_{2}}^{\omega}=0$ if arm $a_{1}$ and arm $a_{2}$
are not connected by connectors (i.e., $\mathbf{K}_{a_{1}}^{a_{2}}=\emptyset$),
which is guaranteed by Eq.(\ref{eq:RoutePlan_6}).

$\bar{\mu}_{a}^{\omega}(t)$ is an auxiliary binary variable for the
convenience of modeling. It is related to $\bar{t}_{a}^{\omega}$
in the following Eq. (\ref{eq:LeaveArm_2}):

\begin{equation}
-\mathrm{M}\left(1-\bar{\mu}_{a}^{\omega}(t)\right)\le t\cdot\Delta t-\bar{t}_{a}^{\omega}\le\mathrm{M}\bar{\mu}_{a}^{\omega}(t),\forall t=0,\dots,T;a\in\mathbf{A};\omega\in\boldsymbol{\Omega}\label{eq:LeaveArm_2}
\end{equation}

\noindent Eq. (\ref{eq:LeaveArm_2}) indicates that $\bar{\mu}_{a}^{\omega}(t)=1$
if $t\cdot\Delta t\ge\bar{t}_{a}^{\omega}$;$\bar{\mu}_{a}^{\omega}(t)=0$,
otherwise.

\subsubsection{No lane changing zone}

If vehicle $\omega$ travels in the connector part of a non-destination
arm within the intersection area or outside the control zone in the
destination arm (i.e., $\bar{\mu}_{a}^{\omega}(t+1)=1$), then vehicle
$\omega$ will beconstrained not to change lanes:

\begin{equation}
\delta_{k}^{\omega}(t+1)-\delta_{k}^{\omega}(t)\le1-\bar{\mu}_{a}^{\omega}(t+1),\forall t=0,\dots,T-1;k\in\mathbf{K}_{a};a\in\mathbf{A}_{0}^{\omega};\omega\in\boldsymbol{\Omega}\label{eq:NoLaneChange}
\end{equation}

\noindent Eq.(\ref{eq:NoLaneChange}) guarantees that $\delta_{k}^{\omega}(t+1)=\delta_{k}^{\omega}(t)$
when $\bar{\mu}_{a}^{\omega}(t+1)=1$.

\subsubsection{Spatial safety gaps}

When two vehicles travel in the same lane in the same arm, a spatial
gap $d$ should be applied for safety concerns:

\begin{equation}
\left|x_{a}^{\omega_{2}}(t)-x_{a}^{\omega_{1}}(t)\right|\ge d-\mathrm{M}\left(1-\rho_{a}^{\omega_{1},\omega_{2}}(t)\right),\forall t=0,\dots,T;a\in\mathbf{A}_{0}^{\omega_{1}}\cap\mathbf{A}_{0}^{\omega_{2}};\omega_{1},\omega_{2}\in\boldsymbol{\Omega}\label{eq:SpatialSafetyGaps_1}
\end{equation}

\noindent where $\rho_{a}^{\omega_{1},\omega_{2}}(t)$ is an auxiliary
binary variable. $\rho_{a}^{\omega_{1},\omega_{2}}(t)=1$ if vehicle
$\omega_{1}$ and vehicle $\omega_{2}$ travel in the same lane (i.e.,
$\sum_{k\in\mathbf{K}_{a}}\left|\delta_{k}^{\omega_{1}}(t)-\delta_{k}^{\omega_{2}}(t)\right|=0$)
in the link part of arm $a$ at time step $t$ (i.e., $\underline{\mu}_{a}^{\omega_{1}}(t)=\underline{\mu}_{a}^{\omega_{2}}(t)=1$
and $\bar{\mu}_{a}^{\omega_{1}}(t)=\bar{\mu}_{a}^{\omega_{2}}(t)=0$).
In that case, Eq. (\ref{eq:SpatialSafetyGaps_1}) is effective. $\rho_{a}^{\omega_{1},\omega_{2}}(t)$
is constrained by

\begin{equation}
\begin{alignedat}{1} & \underline{\mu}_{a}^{\omega_{1}}(t)-\bar{\mu}_{a}^{\omega_{1}}(t)+\underline{\mu}_{a}^{\omega_{2}}(t)-\bar{\mu}_{a}^{\omega_{2}}(t)-\sum_{k\in\mathbf{K}_{a}}\left|\delta_{k}^{\omega_{1}}(t)-\delta_{k}^{\omega_{2}}(t)\right|-1\le\rho_{a}^{\omega_{1},\omega_{2}}(t)\\
 & \forall t=0,\dots,T;a\in\mathbf{A}_{0}^{\omega_{1}}\cap\mathbf{A}_{0}^{\omega_{2}};\omega_{1},\omega_{2}\in\boldsymbol{\Omega}
\end{alignedat}
\label{eq:SpatialSafetyGaps_2}
\end{equation}

\noindent Eq. (\ref{eq:SpatialSafetyGaps_2}) guarantees that $\rho_{a}^{\omega_{1},\omega_{2}}(t)=1$
when $\underline{\mu}_{a}^{\omega_{1}}(t)=\underline{\mu}_{a}^{\omega_{2}}(t)=1$,
$\bar{\mu}_{a}^{\omega_{1}}(t)=\bar{\mu}_{a}^{\omega_{2}}(t)=0$,
and $\sum_{k\in\mathbf{K}_{a}}\left|\delta_{k}^{\omega_{1}}(t)-\delta_{k}^{\omega_{2}}(t)\right|=0.$
Note that $\rho_{a}^{\omega_{1},\omega_{2}}(t)$ will be unconstrained
(i.e., $\rho_{a}^{\omega_{1},\omega_{2}}(t)$ is not necessarily zero)
if vehicle $\omega_{1}$ and vehicle $\omega_{2}$ travel in different
lanes, which can still disable constraints (\ref{eq:SpatialSafetyGaps_1}).

\subsubsection{Temporal safety gaps}

\begin{figure}[tbph]
\noindent \begin{centering}
\includegraphics[width=0.5\linewidth]{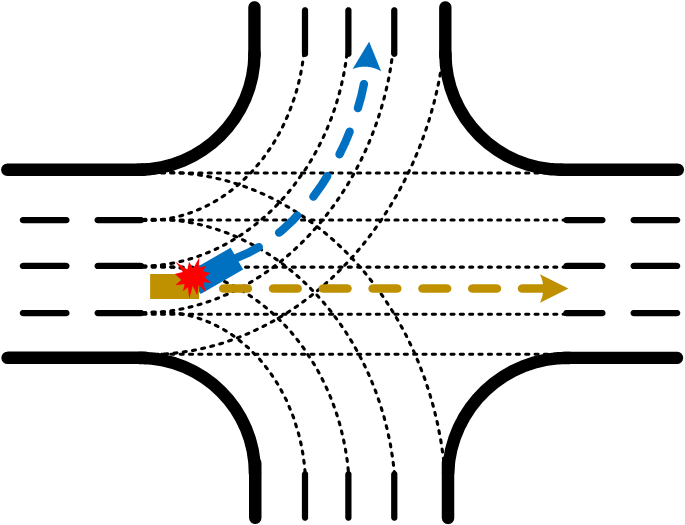}
\par\end{centering}
\caption{Illustration of diverging conflicts.\label{fig:DivergingConflicts}}
\end{figure}

When two vehicles consecutively pass the stop bar in the same lane
in arm $a\ne a_{out}^{\omega}$, a temporal gap $\tau$ is applied
between their passing times for safety concerns:

\begin{equation}
\begin{alignedat}{1} & \left|\bar{t}_{a}^{\omega_{1}}-\bar{t}_{a}^{\omega_{2}}\right|\ge\tau-\mathrm{M}\left(2-\beta_{a}^{\omega_{1}}-\beta_{a}^{\omega_{2}}+\sum_{k\in\mathbf{K}_{a}}\left|\delta_{k}^{\omega_{1}}(T)-\delta_{k}^{\omega_{2}}(T)\right|\right)\\
 & \forall a\in\mathbf{A}_{0}^{\omega_{1}}\cap\mathbf{A}_{0}^{\omega_{2}},a\ne a_{out}^{\omega};\omega_{1},\omega_{2}\in\boldsymbol{\Omega}
\end{alignedat}
\label{eq:TemporalSafetyGaps_1}
\end{equation}
Eq. (\ref{eq:TemporalSafetyGaps_1}) is set to avoid diverging conflicts
as shown in Fig. \ref{fig:DivergingConflicts}. Eq. (\ref{eq:TemporalSafetyGaps_1})
is effective if vehicle $\omega_{1}$ and vehicle $\omega_{2}$ both
plan to visit arm $a$ (i.e., $\beta_{a}^{\omega_{1}}=\beta_{a}^{\omega_{2}}=1$)
and leave the link part in the same lane (i.e., $\sum_{k\in\mathbf{K}_{a}}\left|\delta_{k}^{\omega_{1}}(T)-\delta_{k}^{\omega_{2}}(T)\right|=0$).

\noindent 

\subsubsection{Collision avoidance within intersection areas}

\begin{figure}[tbph]
\noindent \begin{centering}
\includegraphics[width=0.6\linewidth]{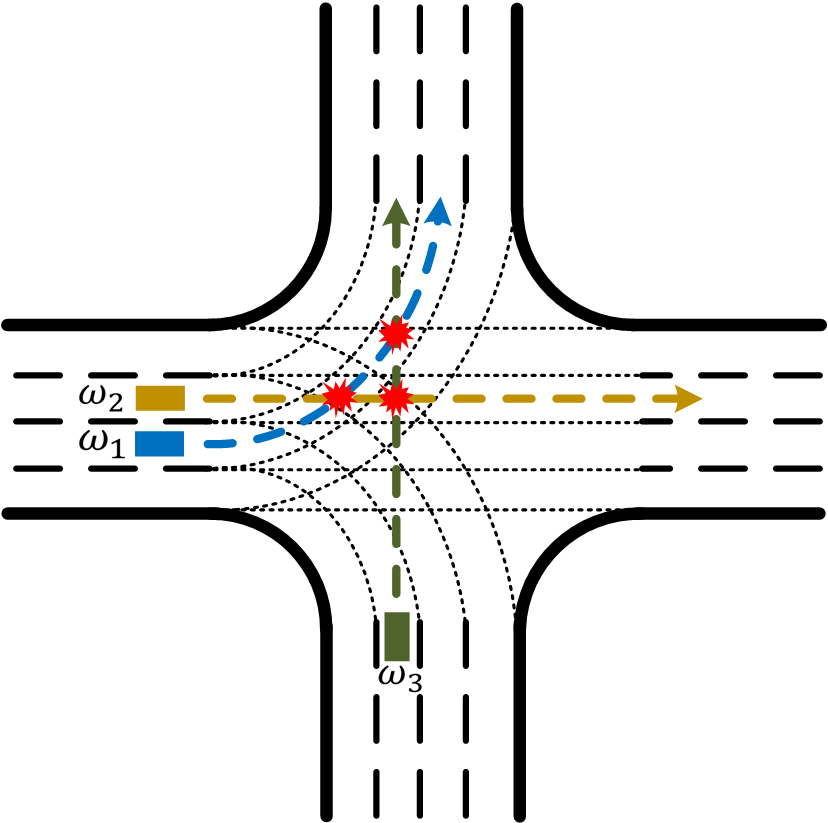}
\par\end{centering}
\caption{Conflicts within the intersection area.\label{fig:IntersectionConflicts}}
\end{figure}

Suppose vehicle $\omega_{1}$ plans to travel from lane $k_{1}$ in
arm $a_{1}$ to lane $k_{2}$ in arm $a_{2}$ via connector $\left\langle k_{1},k_{2}\right\rangle $
(i.e., $\gamma_{a_{1},a_{2}}^{\omega_{1}}=\delta_{k_{1}}^{\omega_{1}}(T)=1$)
and vehicle $\omega_{2}$ plans to travel from lane $k_{3}$ in arm
$a_{3}$ to lane $k_{4}$ in arm $a_{4}$ via connector $\left\langle k_{3},k_{4}\right\rangle $
(i.e., $\gamma_{a_{3},a_{4}}^{\omega_{2}}=\delta_{k_{3}}^{\omega_{2}}(T)=1$)
as shown in Fig. \ref{fig:IntersectionConflicts}. There is a conflict
point between connector $\left\langle k_{1},k_{2}\right\rangle $
and connector $\left\langle k_{3},k_{4}\right\rangle $. For safety
concerns, a temporal gap $\tau$ is applied between their passing
times at the conflict point:

\begin{equation}
\begin{aligned} & \left|\left(\bar{t}_{a_{1}}^{\omega_{1}}+\frac{l_{k_{1},k_{2}}^{p}}{v_{k_{1}}^{k_{2}}}\right)-\left(\bar{t}_{a_{3}}^{\omega_{2}}+\frac{l_{k_{3},k_{4}}^{p}}{v_{k_{3}}^{k_{4}}}\right)\right|\ge\tau-\mathrm{M}\left(4-\gamma_{a_{1},a_{2}}^{\omega_{1}}-\delta_{k_{1}}^{\omega_{1}}(T)-\gamma_{a_{3},a_{4}}^{\omega_{2}}-\delta_{k_{3}}^{\omega_{2}}(T)\right)\\
 & \forall p\in\mathbf{P}_{k_{1},k_{2}}^{k_{3},k_{4}};k_{3}\in\mathbf{K}_{a_{3}}^{a_{4}},k_{4}=k_{3+}^{a_{4}};k_{1}\in\mathbf{K}_{a_{1}}^{a_{2}},k_{2}=k_{1+}^{a_{2}};\\
 & a_{1},a_{2}\in\mathbf{A}_{0}^{\omega_{1}},a_{3},a_{4}\in\mathbf{A}_{0}^{\omega_{2}},a_{4}\ne a_{1},a_{3}\ne a_{2};\omega_{1},\omega_{2}\in\boldsymbol{\Omega}
\end{aligned}
\label{eq:IntersectionConflicts_1}
\end{equation}

\noindent where $\mathbf{P}_{k_{1},k_{2}}^{k_{3},k_{4}}$ is the set
of conflict points between connector $\left\langle k_{1},k_{2}\right\rangle $
and connector $\left\langle k_{3},k_{4}\right\rangle $, which may
have more than one points for a general case; $l_{k_{1},k_{2}}^{p}$
is the distance between the start of connector $\left\langle k_{1},k_{2}\right\rangle $
and conflict point $p$.

\noindent 

\begin{figure}[tbph]
\begin{centering}
\includegraphics[width=0.6\linewidth]{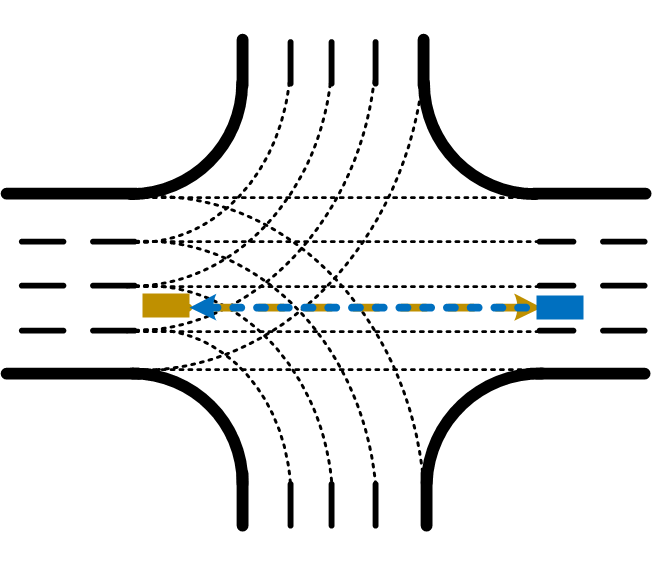}
\par\end{centering}
\caption{Illustration of collision in same connector.\label{fig:Illustration-of-collision-1}}
\end{figure}

Besides Eq. (\ref{eq:IntersectionConflicts_1}), another case needs
special attention. Suppose vehicle $\omega_{1}$ plans to travel from
lane $k_{1}$ in arm $a_{1}$ to lane $k_{2}$ in arm $a_{2}$ via
connector $\left\langle k_{1},k_{2}\right\rangle $ and vehicle $\omega_{2}$
plans to travel from lane $k_{2}$ in arm $a_{2}$ to lane $k_{1}$
in arm $a_{1}$ via connector $\left\langle k_{2},k_{1}\right\rangle $.
In that case, there are countless conflict points in $\mathbf{P}_{k_{1},k_{2}}^{k_{3},k_{4}}$,
which cannot be covered by constraints (\ref{eq:IntersectionConflicts_1})
as shown in Fig.\ref{fig:Illustration-of-collision-1}. The following
Eqs. (\ref{eq:IntersectionConflicts_2}) and (\ref{eq:IntersectionConflicts_3})
are applied instead:

\begin{equation}
\begin{aligned} & \bar{t}_{a_{1}}^{\omega_{1}}-\underline{t}_{a_{1}}^{\omega_{2}}\ge\tau-\mathrm{M}\left(4-\gamma_{a_{1},a_{2}}^{\omega_{1}}-\delta_{k_{1}}^{\omega_{1}}(T)-\gamma_{a_{2},a_{1}}^{\omega_{2}}-\delta_{k_{2}}^{\omega_{2}}(T)+\pi_{k_{1},k_{2}}^{\omega_{1},\omega_{2}}\right)\\
 & \forall k_{1}\in\mathbf{K}_{a_{1}}^{a_{2}},k_{2}=k_{1+}^{a_{2}};a_{1},a_{2}\in\mathbf{A}_{0}^{\omega_{1}}\cap\mathbf{A}_{0}^{\omega_{2}};\omega_{1},\omega_{2}\in\boldsymbol{\Omega}
\end{aligned}
\label{eq:IntersectionConflicts_2}
\end{equation}

\begin{equation}
\begin{aligned} & \bar{t}_{a_{2}}^{\omega_{2}}-\underline{t}_{a_{2}}^{\omega_{1}}\ge\tau-\mathrm{M}\left(5-\gamma_{a_{1},a_{2}}^{\omega_{1}}-\delta_{k_{1}}^{\omega_{1}}(T)-\gamma_{a_{2},a_{1}}^{\omega_{2}}-\delta_{k_{2}}^{\omega_{2}}(T)-\pi_{k_{1},k_{2}}^{\omega_{1},\omega_{2}}\right)\\
 & \forall k_{1}\in\mathbf{K}_{a_{1}}^{a_{2}},k_{2}=k_{1+}^{a_{2}};a_{1},a_{2}\in\mathbf{A}_{0}^{\omega_{1}}\cap\mathbf{A}_{0}^{\omega_{2}};\omega_{1},\omega_{2}\in\boldsymbol{\Omega}
\end{aligned}
\label{eq:IntersectionConflicts_3}
\end{equation}

\noindent where $\pi_{k_{1},k_{2}}^{\omega_{1},\omega_{2}}$ is an
auxiliary binary variable. $\pi_{k_{1},k_{2}}^{\omega_{1},\omega_{2}}=0$
if vehicle $\omega_{1}$ enters connector $\left\langle k_{1},k_{2}\right\rangle $
after vehicle $\omega_{2}$ leaves connector $\left\langle k_{2},k_{1}\right\rangle $
(i.e., $\bar{t}_{a_{1}}^{\omega_{1}}>\underline{t}_{a_{1}}^{\omega_{2}}$);
$\pi_{k_{1},k_{2}}^{\omega_{1},\omega_{2}}=1$, otherwise. Eq. (\ref{eq:IntersectionConflicts_2})
is effective when $\pi_{k_{1},k_{2}}^{\omega_{1},\omega_{2}}=0$ and
Eq. (\ref{eq:IntersectionConflicts_3}) is effective when $\pi_{k_{1},k_{2}}^{\omega_{1},\omega_{2}}=1$.

\subsection{Objective function}

\begin{figure}[tbph]
\noindent \begin{centering}
\includegraphics[width=0.6\linewidth]{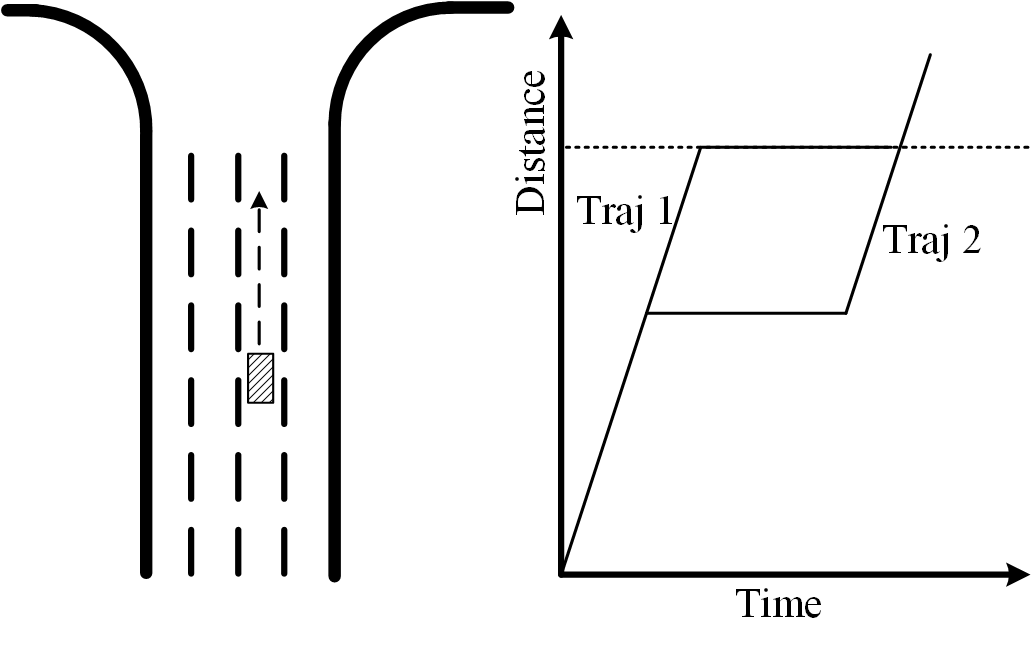}
\par\end{centering}
\caption{Illustration of multiple trajectory solutions.\label{fig:MultipleSolutions}}
\end{figure}

The objective of the optimization model is to minimize total vehicle
delay. Vehicle delay is defined as the difference between actual travel
time and the free-flow travel time. The actual travel time is calculated
as the difference between the times when a vehicle leaves and enters
the control zone. The free-flow travel time is determined from the
movement of each vehicle. Therefore, minimizing vehicle delay is equivalent
to minimizing vehicle's leaving time as the entering time is a constant.
The objective function is formulated as

\begin{equation}
\min\sum_{\omega\in\boldsymbol{\Omega}}\bar{t}_{a_{out}^{\omega}}^{\omega}\label{eq:PrimaryObj}
\end{equation}

\noindent where $\bar{t}_{a_{out}^{\omega}}^{\omega}$ is the time
when vehicle $\omega$ leaves the control zone of link part of the
destination arm $a_{out}^{\omega}$, which means leaving the control
zone. However, multiple optimal trajectory solutions may exist in
terms of total vehicle delay. And the vehicle trajectories of certain
solutions are unfavorable. For example, the two trajectories in Fig.
\ref{fig:MultipleSolutions} have the same delay. But the second trajectory
blocks traffic in the middle of the arm and the first trajectory is
preferred. To this end, a secondary objective is added:

\begin{equation}
\min\sum_{\omega\in\boldsymbol{\Omega}}\sum_{\begin{array}{c}
a\in\mathbf{A}_{0}^{\omega}\\
a\ne a_{out}^{\omega}
\end{array}}\sum_{t=0}^{T}x_{a}^{\omega}(t)\label{eq:SecondaryObj}
\end{equation}

\noindent Objective function (\ref{eq:SecondaryObj}) encourages vehicles
to avoid blocking incoming vehicles in the middle of arms.

\noindent 

To combine Eq. (\ref{eq:PrimaryObj}) and Eq. (\ref{eq:SecondaryObj}),
similar to \citet{Yu2019}, the final objective function is shown
as

\begin{equation}
\min w_{1}\sum_{\omega\in\boldsymbol{\Omega}}\bar{t}_{a_{out}^{\omega}}^{\omega}+w_{2}\sum_{\omega\in\boldsymbol{\Omega}}\sum_{\begin{array}{c}
a\in\mathbf{A}_{0}^{\omega}\\
a\ne a_{out}^{\omega}
\end{array}}\sum_{t=0}^{T}x_{a}^{\omega}(t)
\end{equation}

\noindent where $w_{1}$ and $w_{2}$ are weighting parameters and
$w_{1}\gg w_{2}$ to guarantee the solution quality. Constraints include
Eqs. (\ref{eq:VarDomain_1})–(\ref{eq:IntersectionConflicts_3}).
It is noted that all constraints are linear except Eqs. (\ref{eq:VehDynamics_2}),
(\ref{eq:SpatialSafetyGaps_1}), (\ref{eq:SpatialSafetyGaps_2}),
(\ref{eq:TemporalSafetyGaps_1}), and (\ref{eq:IntersectionConflicts_1})
due to the absolute value function. But they can be easily linearized.
As a result, the proposed model is an MILP model, which can be solved
by many commercial solvers.

\section{Implementation procedure\label{sec:ImpleProcedure}}

The challenge of solving the proposed MILP model lies in the large
dimensions as well as the inclusion of both continuous and binary
variables. Approximately, the number of the variables increases quadratically
with the vehicle number and the connector number. Further, traffic
conditions evolve with new vehicles entering the control zone. The
proposed MILP model needs to be solved to update vehicles' trajectories
considering new vehicles arrivals. Note that the number of the vehicles,
the arms, and the lanes are fixed in each optimization. Then the planning
horizon \textit{$T$} becomes a critical parameter in solving the
proposed model. The model will be infeasible if \textit{$T$} is too
small due to constraint Eq. (\ref{eq:BoundaryCond_3}). However, a
large \textit{$T$} brings intensive computational burden. An algorithm
is designed to adjust \textit{$T$} adaptively and is embedded in
the implementation procedure of the proposed model with varying traffic
conditions:

\noindent \textbf{Step 0}: Initialize the planning horizon $T=T_{0}$
and the simulation time step $t=0$.

\noindent \textbf{Step 1}: Initialize $\mathbf{A}^{\omega}=\mathbf{A}$
for the vehicles that newly enter the control zone.

\noindent \textbf{Step 2}: Update $\mathbf{A}^{\omega}$ for all vehicles
in the control zone as $\mathbf{A}^{\omega}=\mathbf{A}^{\omega}\setminus\left\{ a_{0}^{\omega}\right\} $
if $a_{0}^{\omega}\in\mathbf{A}^{\omega}$.

\noindent \textbf{Step 3}: Get $\mathbf{A}_{0}^{\omega}$ for all
vehicles in the control zone as $\mathbf{A}_{0}^{\omega}=\mathbf{A}^{\omega}\cup\left\{ a_{0}^{\omega}\right\} $.

\noindent \textbf{Step 4}: Collect information from all the vehicles
in the control zone at the current time step $t$.

\noindent \textbf{Step 5}: Solve the MILP model.

\noindent \textbf{Step 6}: If there are no feasible solutions, then
update $T=T+2\Delta T$, where $\Delta T$ is the step length for
adjusting $T$. Go to \textbf{Step 5}. Otherwise, get the solution
$\bar{t}_{a_{out}^{\omega}}^{\omega}$ of each vehicle and go to the
next step.

\noindent \textbf{Step 7}: Update $T=\max\left(\left\lceil \frac{\max_{\omega\in\boldsymbol{\Omega}}\bar{t}_{a_{out}^{\omega}}^{\omega}}{\Delta T}\right\rceil ,T-\Delta T\right)$,
where $\left\lceil \cdot\right\rceil $ is the ceiling function that
maps a real number to the least integer greater than or equal to the
number.

\noindent \textbf{Step 8}: Update the simulation time step $t=t+1$
and go to \textbf{Step 1}.

\section{Numerical studies\label{sec:Numerical-studies}}

\subsection{Experiment design}

To explore the benefits of the proposed LAF control, this study employs
the isolated intersection without lane allocation in Fig. \ref{fig:IntersectionWithRouting}.
Each lane can be used as both approaching and exit lanes for left-turn,
through and right-turn vehicles. Vehicles can take flexible routes
by way of multiple arms to pass through the intersection. The basic
demand of each movement is shown in Table \ref{tab:Basic-traffic-demand},
which is scaled proportionally by a demand factor $\alpha$ as the
input demand. The critical intersection volume-to-capacity (v/c) ratio
\citep{Board2010} of the basic demand is 0.25, which is calculated
as the sum of the critical v/c ratio of each phase with maximum phase
green times. Left-turn, through and right-turn vehicles are taken
into consideration. Low, medium, and high demand levels are tested
with $\alpha=1$, 2, and 4, respectively, which means the v/c ratios
of the low, medium and high demand levels are 0.25, 0.5 and 1 respectively.
The geometric parameters $l_{k_{1}}^{k_{2}}$ and $l_{k_{1},k_{2}}^{p}$
can be easily determined based on the intersection layout.The design
speed $v_{k_{1}}^{k_{2}}$ in a connector is 8 m/s for left-turn vehicles,
10 m/s for through vehicles, and 6 m/s for right-turn vehicles. Other
main parameters are summarized in Table \ref{tab:Main-parameters.}.

\begin{table}[tbph]
\caption{Basic traffic demand.\label{tab:Basic-traffic-demand}}

\noindent \centering{}%
\begin{tabular}{cYYYY}
\hline 
Traffic demand (veh/h) & \multicolumn{4}{c}{To Arm}\tabularnewline
\cline{2-5} \cline{3-5} \cline{4-5} \cline{5-5} 
From Arm & 1 & 2 & 3 & 4\tabularnewline
\hline 
1 & – & 90 & 150 & 30\tabularnewline
2 & 30 & – & 40 & 50\tabularnewline
3 & 150 & 30 & – & 90\tabularnewline
4 & 40 & 50 & 20 & –\tabularnewline
\hline 
\end{tabular}
\end{table}

\begin{table}[tbph]
\caption{Main parameters.\label{tab:Main-parameters.}}

\noindent \centering{}%
\begin{tabular}{YYYYcY}
\hline 
Parameter & Value & Parameter & Value & Parameter & Value\tabularnewline
\hline 
$\Delta t$ & 0.5 s & $T_{0}$ & 30 & $\Delta T$ & 2\tabularnewline
$L_{a}$ & 50 m & $V_{a}$ & 10 m/s & $\tau$ & 1.5 s\tabularnewline
$d$ & 5 m & $w_{1}$ & 300 & $w_{2}$ & 1\tabularnewline
\hline 
\end{tabular}
\end{table}

\begin{figure}[tbph]
\noindent \begin{centering}
\includegraphics[width=0.6\linewidth]{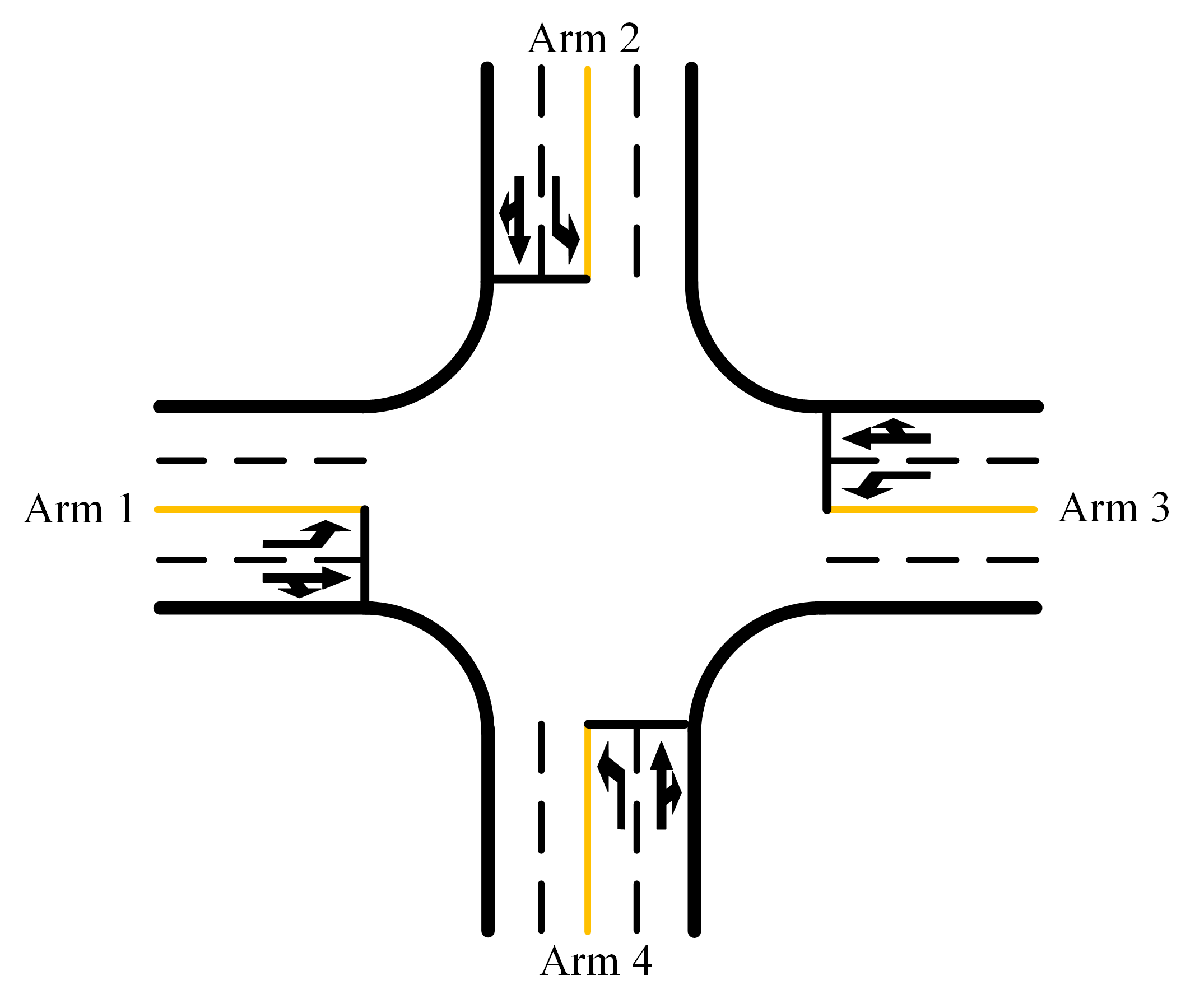}
\par\end{centering}
\caption{Lane allocation in vehicle-actuated control.\label{fig:Lane-allocation}}
\end{figure}

Besides the proposed LAF control, vehicle-actuated control and the
ALAF control in the previous study \citep{Yu2019} are applied as
the benchmarks. In the vehicle-actuated control, the lane allocation
in Fig. \ref{fig:Lane-allocation} and three signal phases are used.
Phase 1 includes the left-turn vehicles in arm 1 and arm 3. Phase
2 includes the through and the right-turn vehicles in arm 1 and arm
3. Phase 3 includes the left-turn, through, and right-turn vehicles
in arm 2 and arm 4.The green extension is 3 s. The all-red clearance
time is 3 s. The minimum green time of each phase is 6 s. The maximum
green times are 15 s, 30 s, and 20 s for phase 1, phase 2, and phase
3, respectively. In the ALAF control, two lanes in each arm are used
for approaching lanes and the remaining two are used for exit lanes.
That is, the approaching lane allocation in Fig. \ref{fig:Lane-allocation}
is removed. The other parameters remain the same as the proposed LAF
control for a fair comparison.

The control algorithms are written in C\#. The LAF control model and
the ALAF control model are solved using Gurobi 8.1.0 (\citet{GurobiOptimization2019}).
The proposed optimization model is executed each time when new vehicles
enter the control zone. The simulation is conducted in SUMO (Simulation
of Urban Mobility) \citep{Krajzewicz2012} on a server with an Intel
2.4 GHz 12-core CPU with 128 GB memory. Only the trajectories of newly
arrived vehicles are optimized for computational efficiency at the
cost of system optimality. The trajectories of the vehicles in the
control zone that have been optimized in the previous optimization
processes are considered in the constraints to avoid collisions. Each
optimization is finished within five minutes. The default lane-changing
and car-following models in SUMO are used in the vehicle-actuated
control. The acceleration/deceleration rates are set as infinity in
SUMO so that vehicles can change speeds and lanes instantaneously
in the benchmark cases for a fair comparison. Five random seeds are
used in the simulation for each demand scenario considering stochastic
vehicle arrivals. Each simulation run is 1200 s with a warm-up period
of 20 s. (\href{https://magic.tongji.edu.cn/en/index.php?catid=41}{Simulation Video})

\subsection{Result and analysis}

To compare the performance of the vehicle-actuated control, the ALAF
control, and the proposed LAF control, average vehicle delay and throughput
are recorded as the performance measures. The delay of a vehicle is
calculated as the difference between the actual travel time and the
free-flow travel time of its movement. Only the delays of the vehicles
that have left the control zone are counted. The simulation results
are shown in Table \ref{tab:Average-vehicle-delay} and Table \ref{tab:throughput}.

\begin{table}[tbph]
\caption{Average vehicle delay (s).\label{tab:Average-vehicle-delay}}

\noindent \centering{}%
\begin{tabular}{c>{\centering}m{0.18\linewidth}>{\centering}m{0.15\linewidth}>{\centering}m{0.28\linewidth}}
\hline 
Average vehicle delay & \multicolumn{3}{c}{Demand Scenarios}\tabularnewline
\cline{2-4} \cline{3-4} \cline{4-4} 
(Standard deviation) & Low ($\alpha=1$) & Medium ($\alpha=2$) & High ($\alpha=4$)\tabularnewline
\hline 
Vehicle-actuated Control & 14.50 (3.49) & 23.13 (1.95) & 34.66 (9.86)\tabularnewline
ALAF Control & 0.71 (0.09 ) & 0.83 (0.08) & 0.91 (0.15 )\tabularnewline
LAF Control & 0.09 (0.07 ) & 0.15 (0.06) & 0.28 (0.06)\tabularnewline
\hline 
\end{tabular}
\end{table}

\begin{table}[tbph]
\caption{Vehicle throughput (veh/h).\label{tab:throughput}}

\noindent \centering{}%
\begin{tabular}{c>{\centering}m{0.18\linewidth}>{\centering}m{0.18\linewidth}>{\centering}m{0.28\linewidth}}
\hline 
Throughput & \multicolumn{3}{c}{Demand Scenarios}\tabularnewline
\cline{2-4} \cline{3-4} \cline{4-4} 
(Standard deviation) & Low ($\alpha=1$) & Medium ($\alpha=2$) & High ($\alpha=4$)\tabularnewline
\hline 
Vehicle-actuated Control & 771 (24–) & 1524 (67–) & 2638 (110)\tabularnewline
ALAF Control & 772 (27) & 1543 (28) & 3069 (36)\tabularnewline
LAF Control & 778 (18) & 1563 (31) & 3096 (42)\tabularnewline
\hline 
\end{tabular}
\end{table}

Table \ref{tab:Average-vehicle-delay} shows that the delays increase
with the demand when the vehicle-actuated control, the ALAF control,
and the LAF control are applied. The average vehicle delay in the
vehicle-actuated control rises more noticeably than those in the ALAF
control and the LAF control when the demand increases from the low
level to the high level. The increased average vehicle delay in the
vehicle-actuated control reaches 20.16 s while the values are only
0.20 s and 0.19 s in the ALAF control and the LAF control, respectively.
Further, the ALAF control and the LAF control significantly outperform
the vehicle-actuated control in terms of average vehicle delays at
all the demand levels. Compared with the vehicle-actuated control,
the ALAF control and the LAF control reduce the average vehicle delays
by more than 90\%, which validates the benefits of the CAV-based intersection
control without lane allocation. It is also observed that the average
vehicle delay in the proposed LAF control are less than one third
of the average vehicle delay in the ALAF control.That is, the proposed
LAF control remarkably outperforms the ALAF control in terms of the
average vehicle delay.

Table \ref{tab:throughput} shows the vehicle throughput of the vehicle-actuated
control, the ALAF control, and the LAF control. At the low and medium
demand levels, the throughput of the three control modes differs insignificantly.
That means the demands are below the intersection capacity. However,
the ALAF control and the LAF control have much higher throughput by
$\sim$17\% at the high demand level. Because the demand exceeds the
intersection capacity in the vehicle-actuated control but is well
accommodated in the ALAF control and the LAF control due to the significantly
improved capacity. The difference between the throughput of the ALAF
control and LAF control is insignificantly. Throughput tests of higher
traffic demands are tested in following part.

\subsection{Sensitivity analysis}

\subsubsection{Demand structures}

\begin{figure}[tbph]
\begin{centering}
\subfloat[\label{fig:Delay-of-ALAF1}]{\begin{centering}
\includegraphics[width=0.5\linewidth]{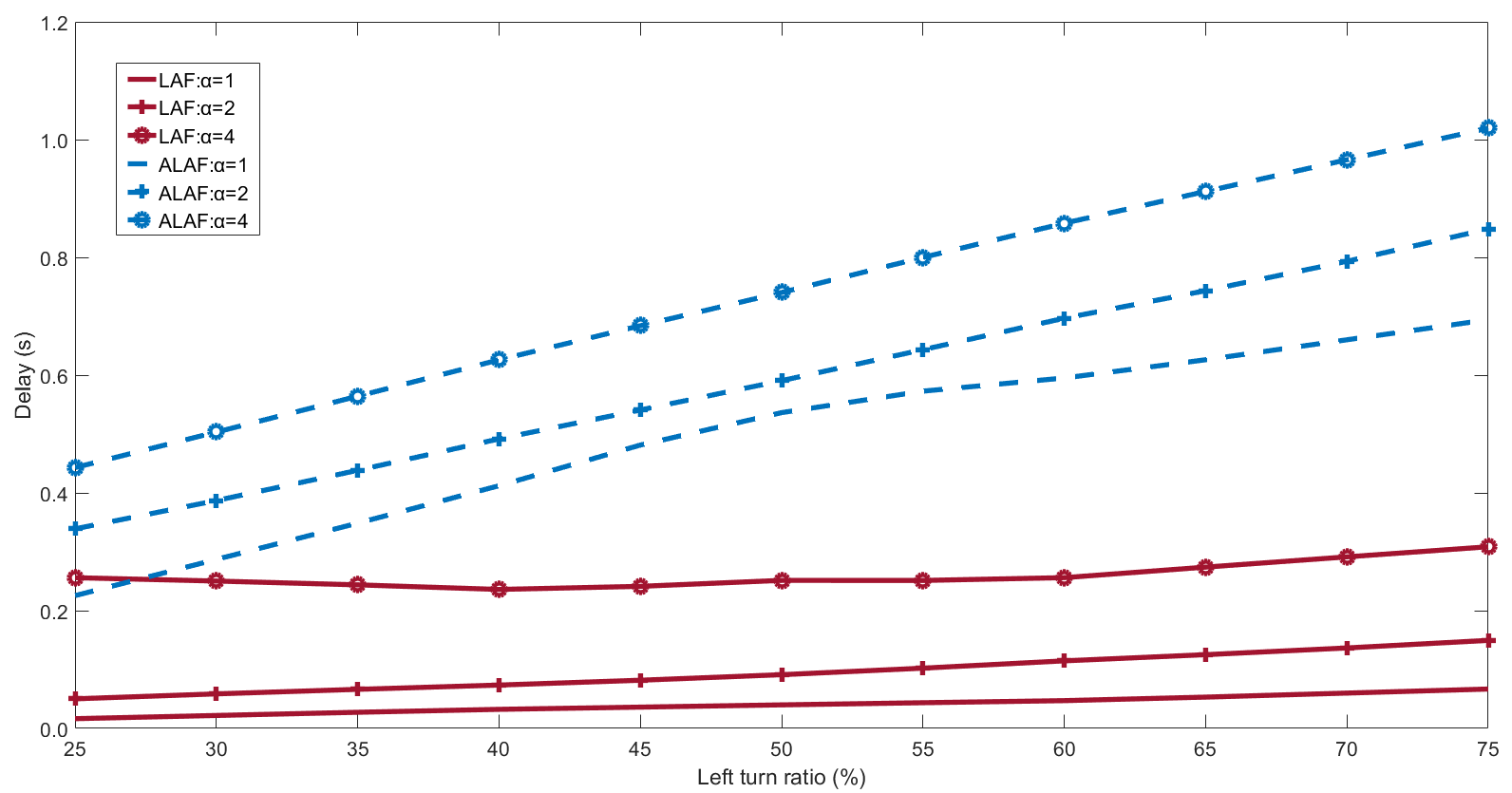}
\par\end{centering}
}\subfloat[\label{fig:Decrease-of-Delay}]{\begin{centering}
\includegraphics[width=0.5\linewidth]{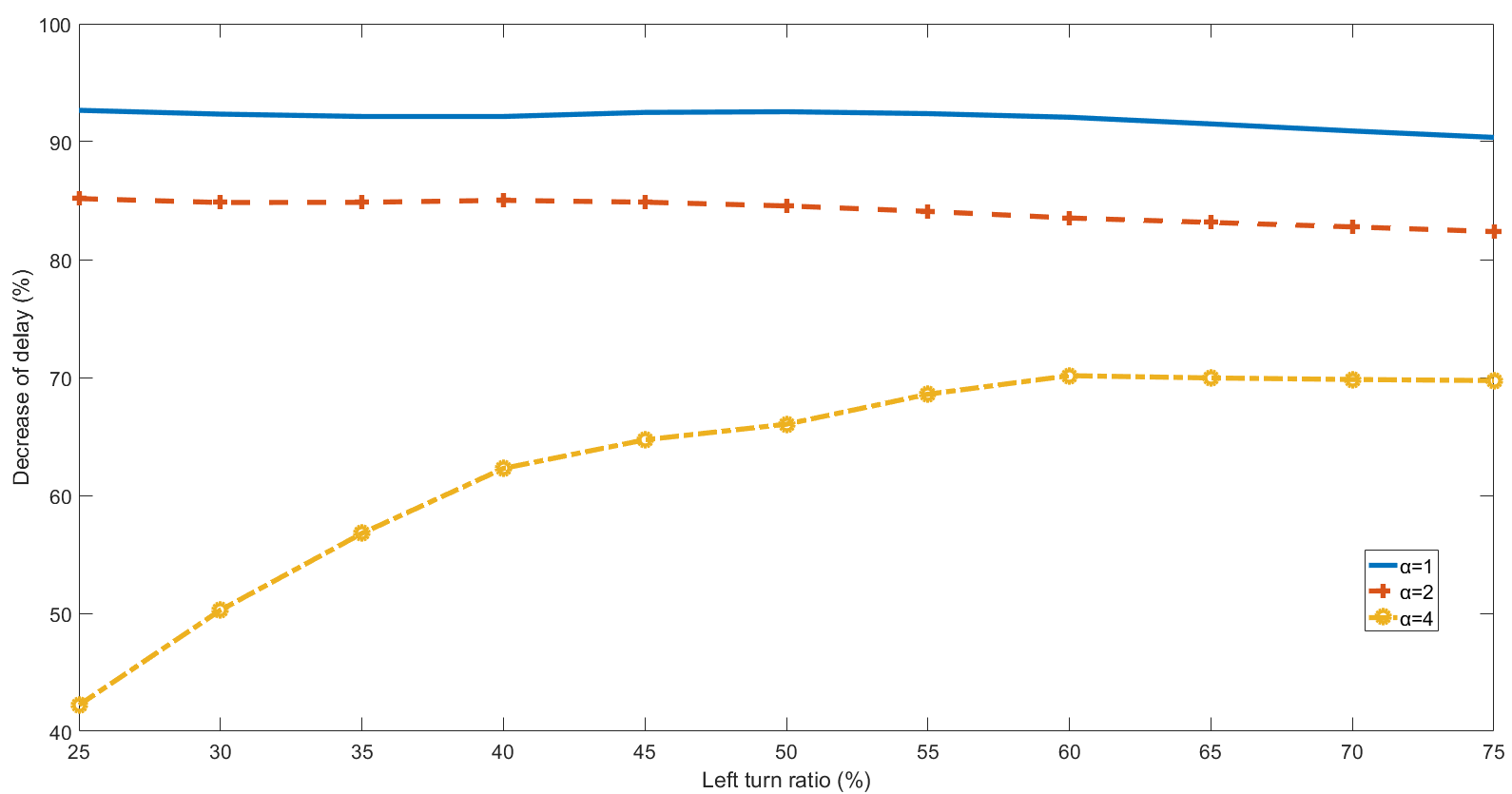}
\par\end{centering}
}
\par\end{centering}
\caption{Simulation results of delay: (a) Delays of ALAF control and LAF control;
(b) Decrease of Delay by using LAF.}
\end{figure}

The delays of the LAF control and the ALAF control with different
demand structures are tested. The results are shown in Fig. \ref{fig:Delay-of-ALAF1}.
The delay of the ALAF control will increase with the growth of the
traffic demand and let-turning ratio. In contrast, the delay of the
LAF control is not sensitive to the left-turn ratio. That is because
that the left-turn vehicles have a shorter shortest path and less
conflict points on the path as shown as the vehicle $\omega_{1}$
in Fig. \ref{fig:IntersectionWithRouting}. In other words, the left-turn
vehicles can be equivalent to right-turn vehicles under the LAF control.
The decreases of delays by using the LAF control instead of the ALAF
control under different demand structures are shown as Fig. \ref{fig:Decrease-of-Delay}.
Under all tested demand structures, at least 40 percent delay can
be saved. The decrease of delay is dropped with the increase of traffic
demand. When traffic demands are low and medium level, more than 90
and 80 percent delay can be saved, respectively. In contrast, the
decrease delays are lower than 70 percent under the high level traffic
demand. As for the influence of the left-turn ratio, the decrease
of delays is not sensitive to the left-turn ratio when traffic demand
is low level or medium level. On the contrary, the decrease of delay
is growth from 40 percent to 70 percent with the increase of the left-turn
ratio when the traffic demand is high level. The LAF control is more
suitable to be used in intersections with large traffic demand and
a large left-turn ratio or intersections with low traffic demand.

\begin{figure}[tbph]
\begin{centering}
\subfloat[\label{fig:Throughput-of-LAF}]{\begin{centering}
\includegraphics[width=0.5\linewidth]{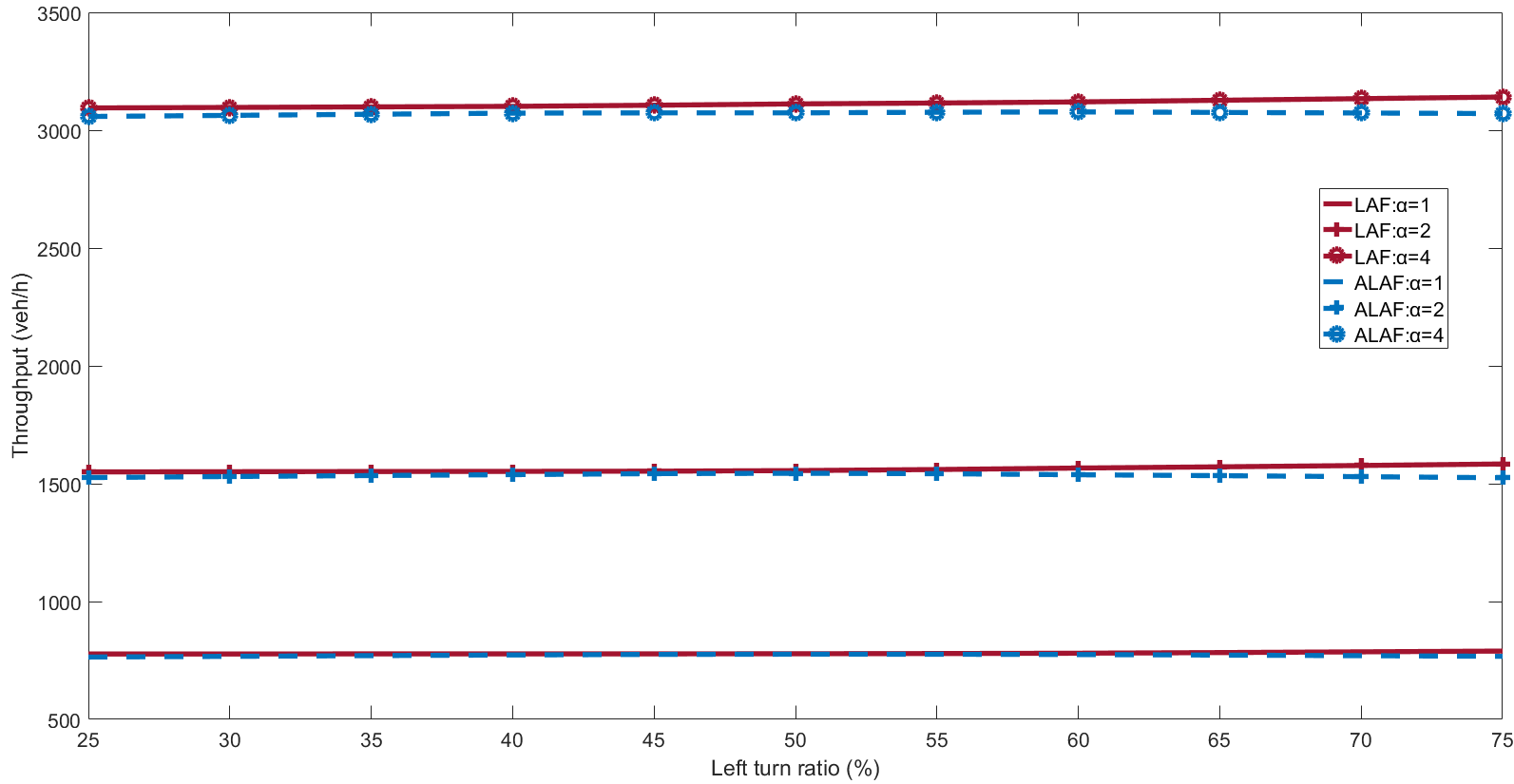}
\par\end{centering}
}\subfloat[\label{fig:Improvement-of-Throughput-1}]{\begin{centering}
\includegraphics[width=0.5\linewidth]{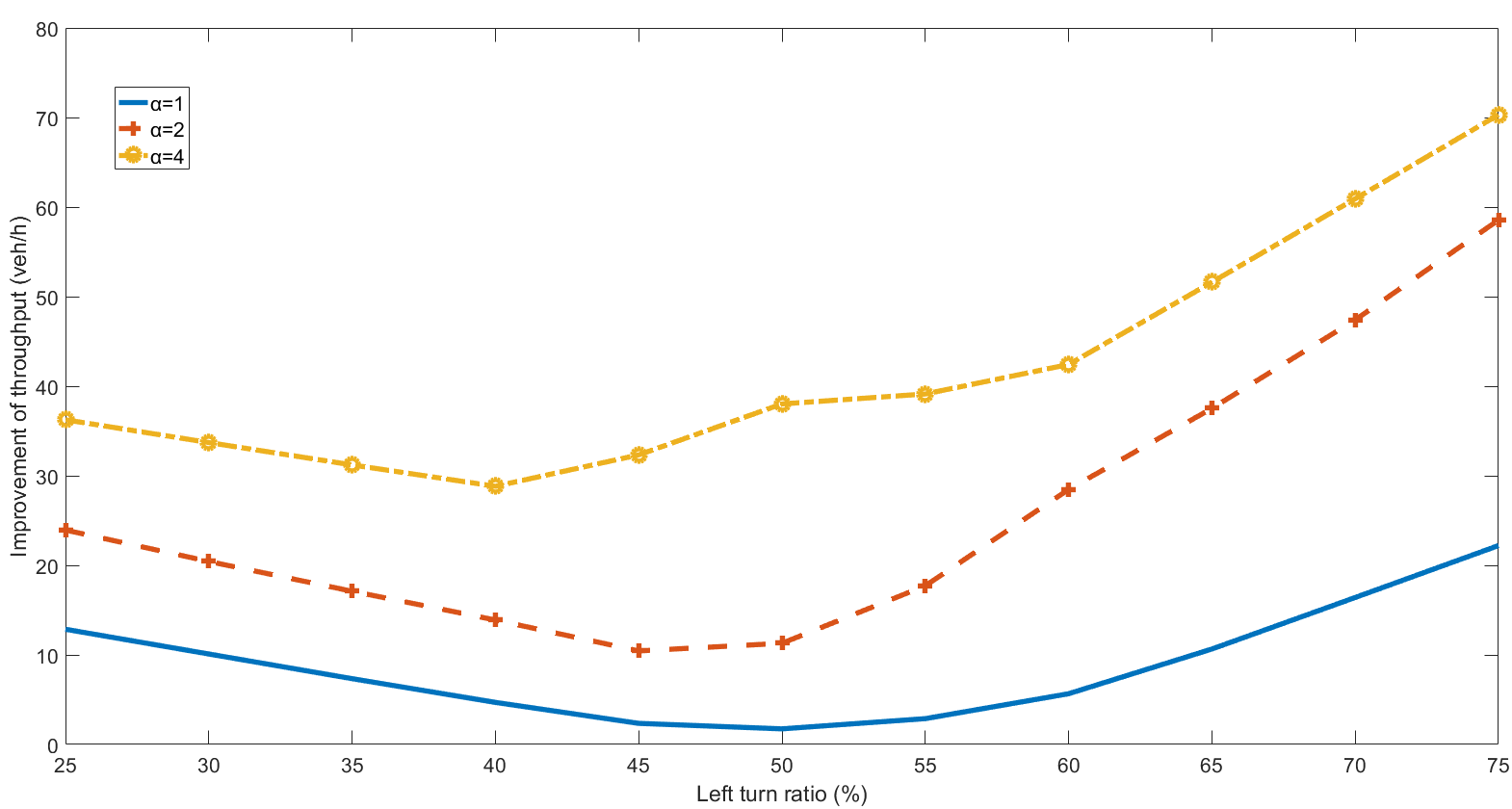}
\par\end{centering}
}
\par\end{centering}
\caption{Improvement of Throughput by using LAF: (a) Throughput of ALAF and
LAF; (b) Improvement of Throughput by using LAF.\label{fig:Improvement-of-Throughput}}
\end{figure}

However, the throughput of the LAF control and the ALAF control are
still very similar with each other under different traffic condition
as shown in \ref{fig:Throughput-of-LAF}. Since the tested demands
are lower than capacity, the throughput is close to tested demands.
And the throughput is not changed with the increase of left-turn ratio.
The comparison of throughput under the LAF control and the ALAF control
is shown in Fig.\ref{fig:Improvement-of-Throughput-1}. The gap between
the LAF control and the ALAF control becomes larger with the growth
of the traffic demand. But the throughput difference between two control
methods can almost be neglected compared with the throughput which
is changing from 770 veh/h to 3080 veh/h.

\begin{figure}[tbph]
\begin{centering}
\includegraphics[width=0.75\linewidth]{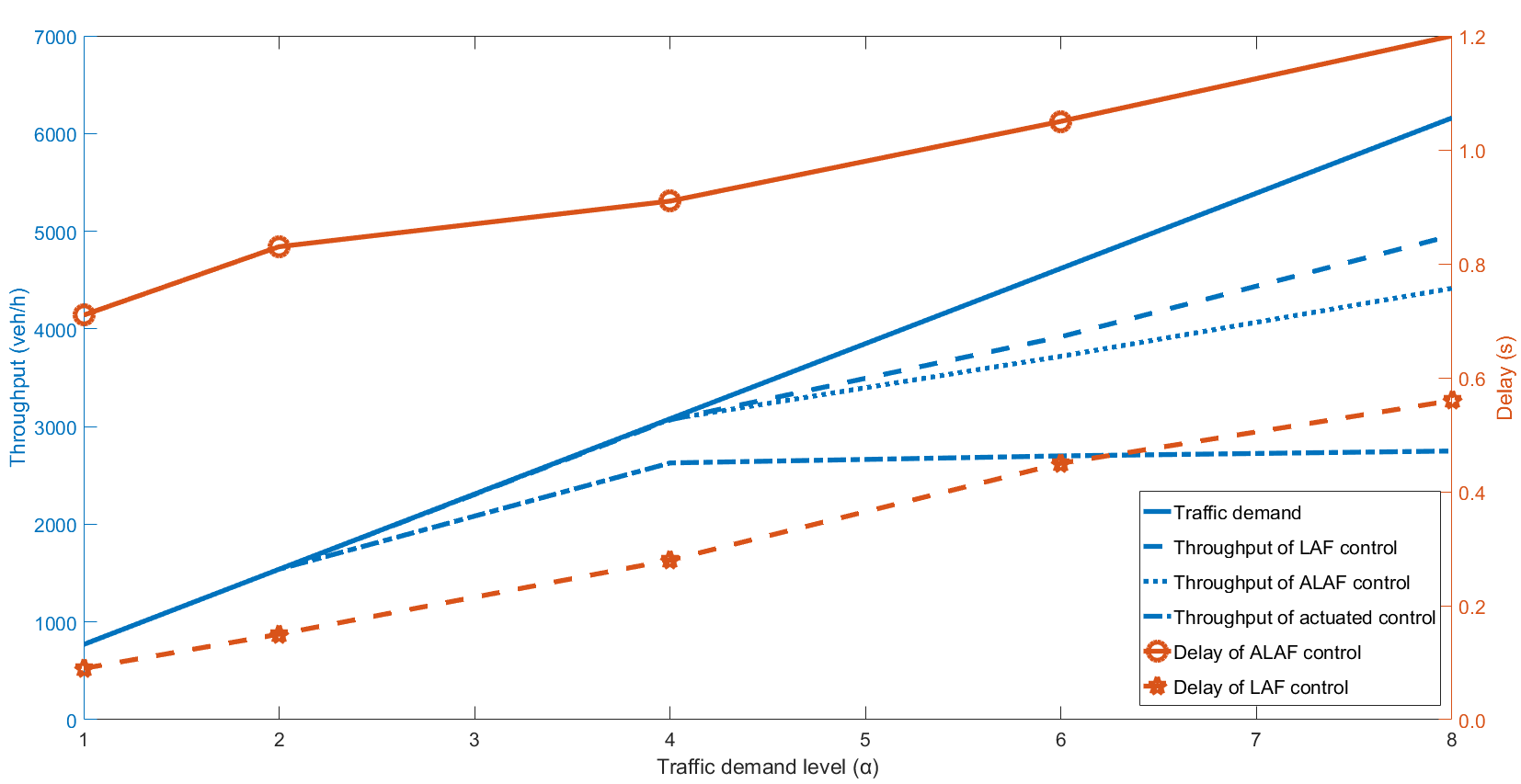}
\par\end{centering}
\caption{Simulation results: vehicle throughput.\label{fig:Simulation-results:-vehicle-throughput}}
\end{figure}

For better exploring the improvement of capacity by using the LAF
control and the ALAF control, higher demands are tested. The result
is shown in Fig.\ref{fig:Simulation-results:-vehicle-throughput}.
Since the gap between the delay of the vehicle-actuated control and
the delay of the LAF/ALAF control is large, the delay of the vehicle-actuated
control is not shown in this figure for a better observation of the
delays of the LAF control and the ALAF control. Under any demand level,
the LAF control has a lower delay than the ALAF control.

As for the throughput, the throughput of the vehicle-actuated control
is hard to growth with the increase of the traffic demand after exceeding
2700 veh/h. In contrast, this situation of the ALAF control and the
LAF control occurred at the value of around 3800 veh/h and 4200 veh/h,
respectively. The capacity is improved after applying the LAF control
and can reach 4200 veh/h. The ALAF control can also improve the capacity
but is less significant than LAF control. The LAF control has better
performance than the ALAF control not only on the delay but also on
the throughput.

The advantages mainly come from two factors: 1) The relaxed constraints
of defining approaching and exit lanes. Each lane can be used as both
approaching and exit lane as long as safety is guaranteed. As a result,
the spatial resources at the intersection can be utilized in a more
effective way. 2) Flexible routing. Vehicles will take a detour by
way of multiple arms to pass through the intersection if less delay
can be achieved. In this way, the solution space of the vehicle trajectory
planning is enlarged and potential better solutions are expected.

\subsubsection{Temporal safety gaps}

\begin{figure}[H]
\begin{centering}
\includegraphics[width=0.75\linewidth]{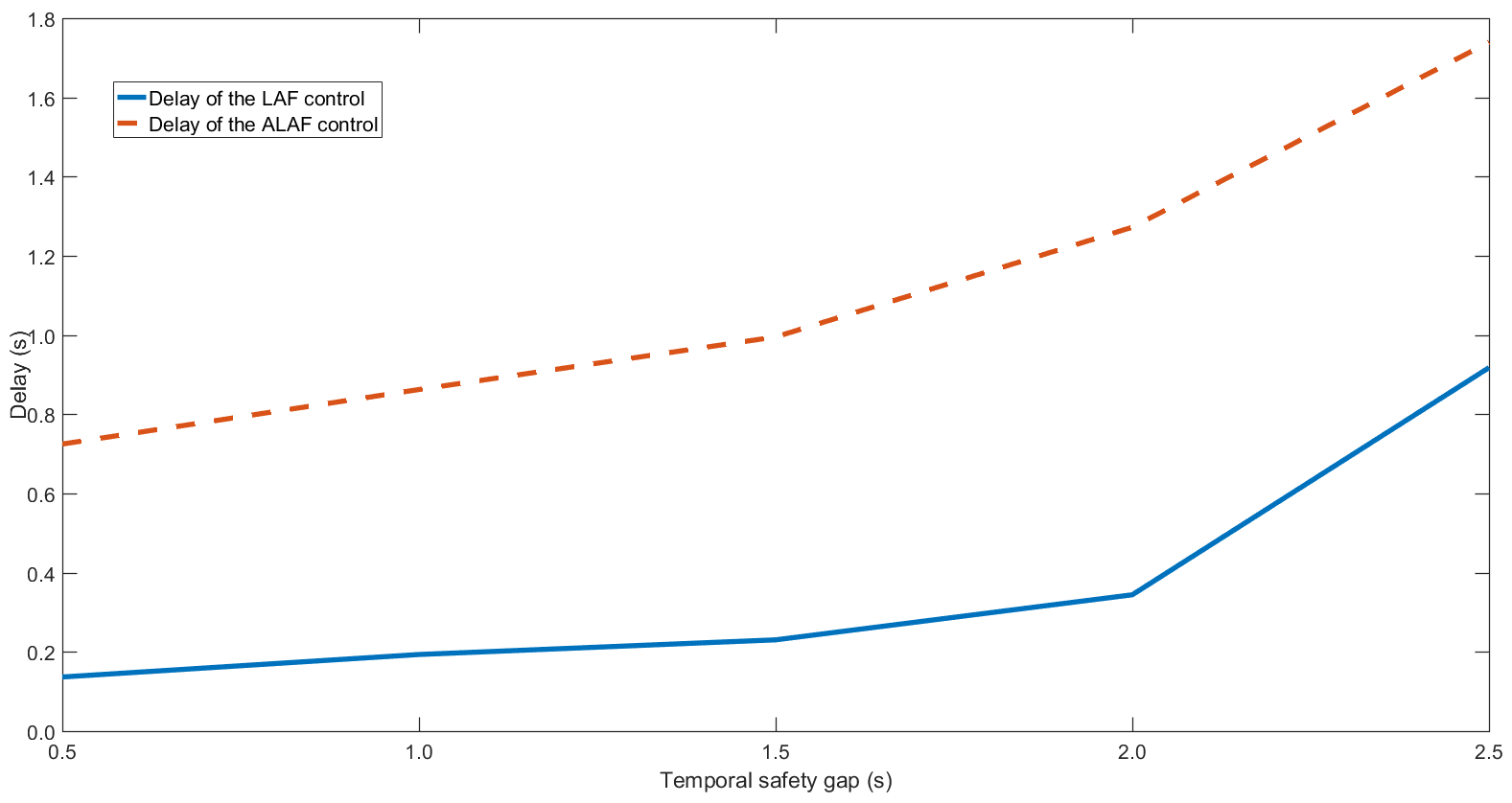}
\par\end{centering}
\caption{Sensitivity analysis of temporal safety gaps.\label{fig:Sensitive_safetygaps}}
\end{figure}

The signal-free management method is used in the LAF control and the
ALAF control. It means the temporal safety gap is a critical parameter.
Because temporal safety gap can influence control efficiency and safety
significantly. Smaller temporal safety gaps may lead to a lower delay,
but result in safety problems. The influence of temporal safety gaps
on the performance of the LAF control and the ALAF control is investigated.
0.5 s is the smallest safety gap for constant time headway policy
of all connected and automated vehicles environment and is selected
as basic parameter by \citet{Bian2019} in their research. As for
regular vehicles, the time headway will increase with the decrease
of speed. The time headway is distributed centered on 2.5 s when the
speed is between 3 m/s and 5 m/s (\citet{Li2010}). Since the smallest
intersection passing speed in this case is 6 m/s which is larger than
5 m/s, the most common time headway should be lower than 2.5 s when
the vehicles are regular vehicles. So the different values of temporal
safety gaps are tested from 0.5 s to 2.5 s per 0.5 s. The result is
shown in Fig. \ref{fig:Sensitive_safetygaps}. The delays under both
control methods increase with the growth of temporal safety gaps.
The delay of the ALAF control is larger than the delay of the LAF
control with any temporal safety gap. Except that, the ALAF control
also has higher rising speed when the temporal safety gap is lower
than 2 s. In other words, the ALAF control is more sensitive to the
change of the temporal safety gap than the LAF control when time headway
is lower than 2 s.

\section{Conclusions and recommendations\label{sec:Conclusions}}

This paper proposes an MILP model to optimize vehicle trajectories
at a ``signal-free'' intersection without lane allocation under
the fully CAV environment. Each lane can be used as both approaching
and exit lanes for all vehicle movements including left-turn, through,
and right-turn. Vehicle can take flexible routes by way of multiple
arms to pass throughput the intersection. The interactions between
vehicle trajectories are modeled explicitly at the microscopic level.
Car-following and lane-changing behaviors of the vehicles within the
control zone can be optimized in one unified framework in terms of
total vehicle delay. In the implementation procedure, the planning
horizon is adaptively adjusted to make a balance between the feasibility
of the MILP model and computational efficiency. In the numerical studies,
only the trajectories of newly arrived vehicles are optimized for
computational efficiency at the cost of system optimality. The simulation
results show that the proposed LAF control outperforms the vehicle-actuated
control and the ALAF control in the previous study \citep{Yu2019}
in terms of both vehicle delay and throughput. The sensitivity analysis
further validates the advantages of the LAF control over the ALAF
control with different demand structures and temporal safety gaps.

This study assumes a fully CAV environment. However, regular vehicles,
CVs, and CAVs will coexist in the near future. It is worthwhile to
investigate the control methods under the mixed traffic environment.
This study focuses on isolated intersections. It is planned to extend
the proposed model to a corridor and a network. For simplicity, the
first-order vehicle dynamics models are used in this paper. It is
not difficult to apply higher-order vehicle dynamics models but the
model will be no longer linear. The solving algorithms could be a
great challenge. The computational burden is heavy due to the large
dimensions of the model, especially, when the trajectories of all
vehicles are optimized at the same time. Efficient algorithms are
expected to balance the solution quality and the computational time.
Issues such as communication delays and detection issues may be inevitable
even when 100\% CAVs are deployed. Robust planning of vehicle trajectories
is another research direction.

\section*{Acknowledgments}

This research was funded by Shanghai Sailing Program (No. 19YF1451600),
the National Natural Science Foundation of China (No. 51722809 and
No. 61773293), and the Fok Ying Tong Education Foundation (No. 151076).
The views presented in this paper are those of the authors alone.

\bibliographystyle{plainnat}
\bibliography{Managing_CAVs_with_flexible_routing_at_isolated_intersections}

\end{document}